\documentclass[twocolumn,preprintnumbers,nofootinbib]{revtex4}
\usepackage{hyperref}
\usepackage{bm}

\usepackage{amsmath}
\usepackage{amsfonts}
\usepackage{slashed}
\usepackage{amssymb}
\usepackage{graphicx}
\usepackage{mathrsfs}
\usepackage{mathtools}
 \usepackage{exscale,relsize}
 
\def\beq{\begin{equation}}
\def\eeq{\end{equation}}
\def\bea{\begin{eqnarray}}
\def\eea{\end{eqnarray}}

\def\d{{\rm d}}
\def\B{{\cal B}}
\def\E{{\cal E}}
\def\0{{\mathbf 0}}

\def\n{{\mathbf{n}}}

\def\x{{\mathbf{x}}}

\RequirePackage{color}

\begin{document}

\title{The Superhorizon Test of Future B-mode Experiments}

\author{Hayden Lee}
\affiliation{Department of Applied Mathematics and Theoretical Physics, Cambridge University, Cambridge, CB3 0WA, UK}
\author{S.-C.~Su}
\affiliation{Department of Applied Mathematics and Theoretical Physics, Cambridge University, Cambridge, CB3 0WA, UK}
\author{Daniel Baumann}
\affiliation{Department of Applied Mathematics and Theoretical Physics, Cambridge University, Cambridge, CB3 0WA, UK}

\begin{abstract}
Inflation predicts B-mode polarization with correlations that span superhorizon scales at recombination. In contrast, the correlations set up by causal sources, such as phase transitions or defects, necessarily vanish on superhorizon scales. Motivated by BICEP2's B-mode detection, we consider the prospects for measuring the inflationary superhorizon 
signature in future observations. We explain that the finite resolution of an experiment and the filtering of the raw data induces a transfer of spurious subhorizon power to superhorizon scales, and describe ways to correct for it.
We also provide a detailed treatment of possible sources of noise in the measurement.
Finally, we present forecasts for the detectability of the signal with future CMB polarization experiments.

\bigskip
\end{abstract}

\maketitle

\section{Introduction}\label{sec:intro}

Two questions have been on every cosmologist's mind ever since the BICEP2 collaboration announced the detection of B-mode polarization~\cite{Ade:2014xna}: Is the signal cosmological? And, is it from inflation? Measuring the precise shape of the B-mode spectrum can help to address both of these questions. 

A distinguished feature of inflationary perturbations is the fact that they are correlated over apparently acausal scales. 
For scalar perturbations, this leads to a distinctive cross-correlation in the cosmic microwave background~(CMB) between temperature perturbations and E-mode polarization~\cite{Coulson:1994qw}.
The detection of superhorizon TE correlations by WMAP~\cite{Peiris:2003ff} is arguably the most convincing piece of evidence that the observed 
density perturbations were generated during inflation~\cite{Spergel:1997vq, Dodelson:2003ip}.  In~\cite{Baumann:2009mq}, it was pointed out that an analogous causality test can be performed for inflationary tensor modes. 
In this paper, we revisit and refine this proposal in light of the BICEP2 result.

\begin{figure}[t!]
\centering
\includegraphics[width=0.47\textwidth]{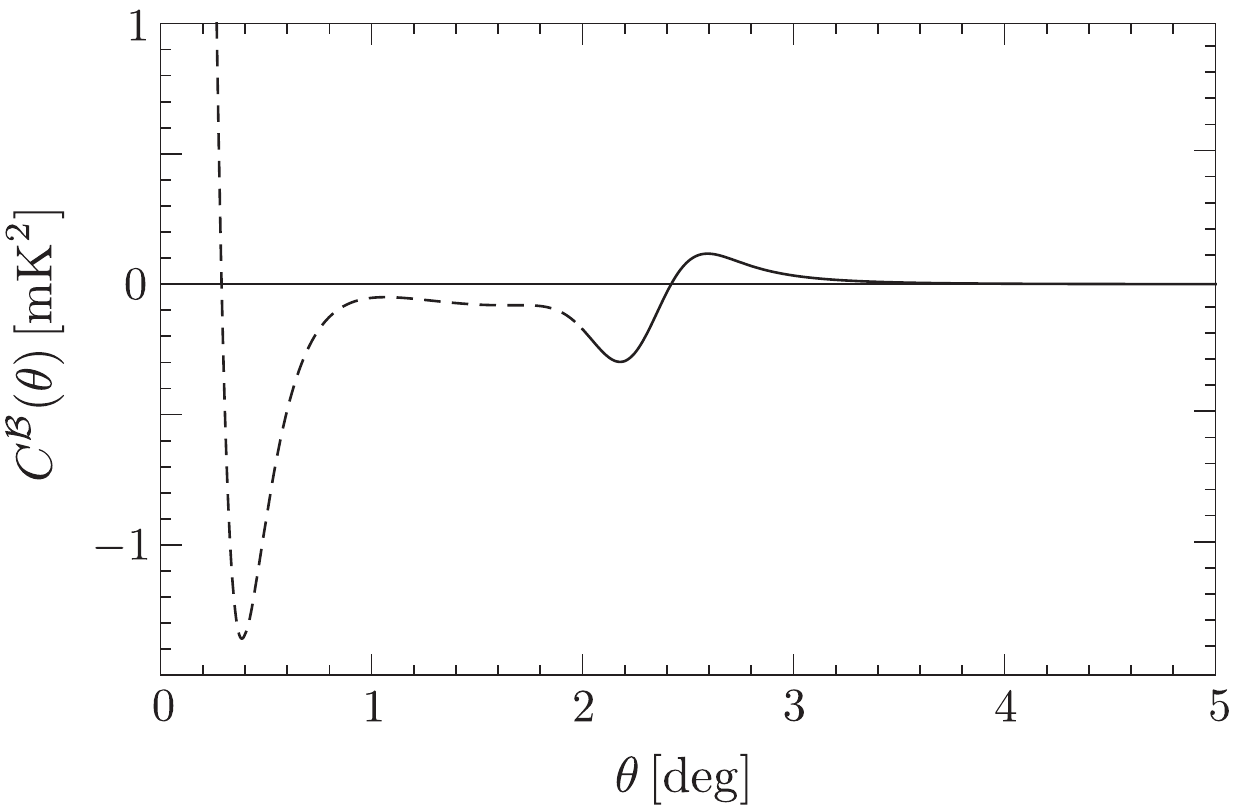}
\vskip -3pt
\caption{\label{fig:ClBtildeAll} Local $\B$-mode correlation function for $r=0.13$.  The dashed and solid parts of the curve represent subhorizon and superhorizon scales, respectively.} 
\label{fig:real}
\end{figure}

The conventional $E$- and $B$-modes~\cite{Kamionkowski:1996ks, Zaldarriaga:1996xe}
are ill-suited to address questions of causality, since they are defined non-locally in terms of the Stokes parameters of the radiation field. We will therefore work with a local alternative~\cite{Smith:2006vq} to the standard $E$- and $B$-modes which we will denote by $\E$ and $\B$.
In the flat-sky limit, we have $\E = \nabla^2 E$ and $\B = \nabla^2 B$, where $\nabla^2$ is the two-dimensional Laplacian in the plane orthogonal to the line-of-sight. 
Fig.~\ref{fig:real} shows inflation's prediction for the $\B$-mode correlation function in real space. Unlike the correlation functions sourced by scalar fluctuations, the signal does not have a peak at the acoustic scale ($2\theta_a \sim 1.2^\circ$). Instead the tensor-induced signal peaks around the horizon scale ($\theta_c \equiv 2\theta_h\sim 2.3^\circ$) corresponding to the time when the inflationary gravitational waves re-entered the horizon and started oscillating. Causality forbids such a superhorizon signal for any post-inflationary mechanism, such as phase transitions~\cite{JonesSmith:2007ne} or defects~\cite{Seljak:1997ii, Pogosian:2007gi}, for gravitational wave production~\cite{Hu:1997hp, Spergel:1997vq}.
A measurement of $\B$-mode correlations above 2 degrees therefore constitutes an important test for the inflationary origin of the signal.

\begin{figure}[t]
\centering
\includegraphics[width=0.47\textwidth]{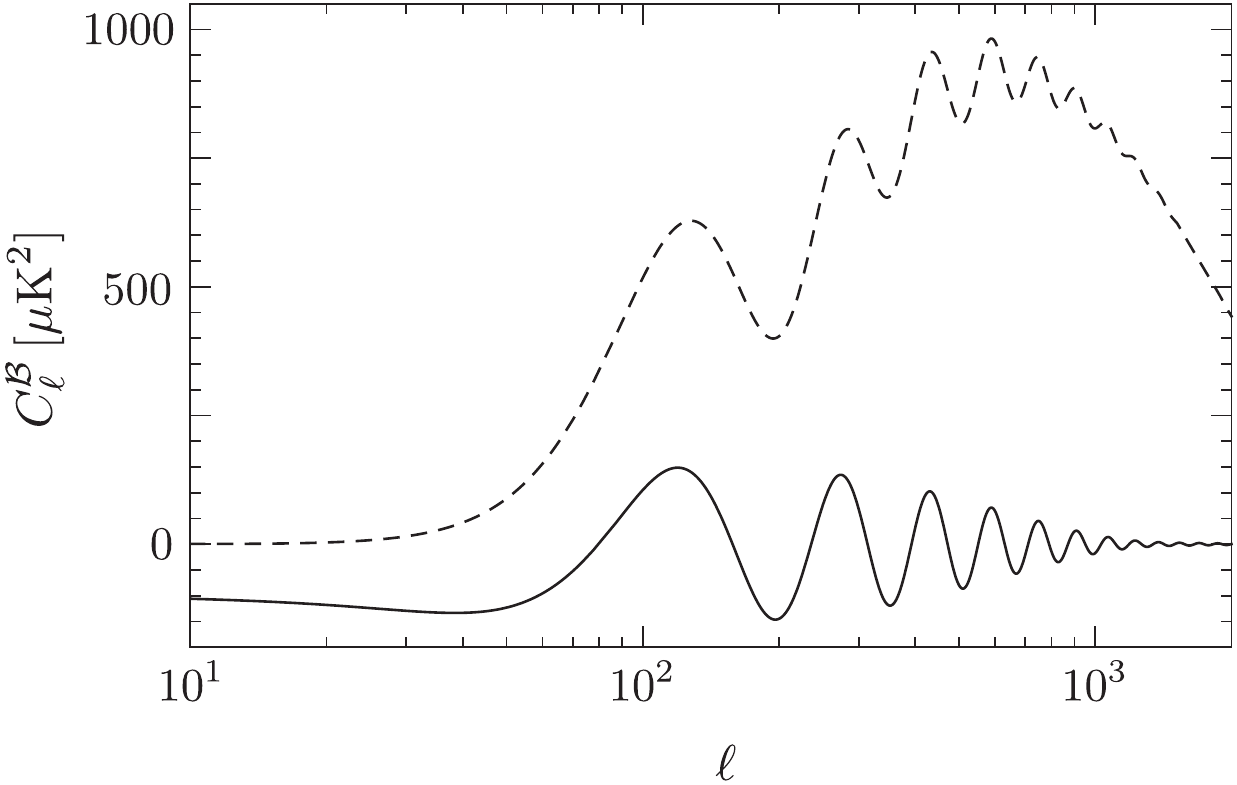}
\vskip -4pt
\caption{\label{fig:powerspectrum}The superhorizon $\B$-mode power spectrum (solid) and the full $\B$-mode power spectrum (dashed) for $r=0.13$.}
\end{figure}

Fig.~\ref{fig:powerspectrum} shows the corresponding superhorizon signal in harmonic space.
(See \S\ref{sec:supsignal} for the precise definition of the superhorizon power spectrum.)
We see that the superhorizon information is encoded in the precise shape and the locations of the peaks of the spectrum.   Notice that the superhorizon signal isn't just in the lowest multipole moments. In fact,
the asymptotic scaling of the spectrum, $C_\ell^\B \sim \ell^4$ for $\ell \ll 80$, is universal and does not help to distinguish causal sources from inflation~\cite{Baumann:2009mq}.

Extra care must be taken when working with $\E$- and $\B$-modes as we are dealing with derivatives of the raw data, corresponding to a blue noise spectrum in harmonic space. 
In order not to become dominated by small-scale noise, smoothing needs to be applied to the data. However, if the smoothing scale is chosen to be too large, it induces a transfer of spurious subhorizon power to superhorizon scales. Conversely, a small smoothing scale reduces the signal-to-noise. We will discuss the optimal strategy for minimizing the effects of spurious modes while maximizing the signal-to-noise for the true superhorizon signal.

\vskip 4pt 
The paper is organized as follows. In \S\ref{sec:signal}, we review the concept of the local $\B$-modes and present the superhorizon $\B$-mode signal predicted by inflation.
In \S\ref{sec:noise}, we examine all potential sources of noise. We show that subhorizon modes can contaminate the superhorizon signal, especially if smoothing is applied to the data  to suppress small-scale noise. In \S\ref{sec:method}, we introduce an estimator of the superhorizon part of the signal and define the signal-to-noise ratio. We also present a measure for the amount of contamination from spurious subhorizon modes and describe ways to minimize their effects. 
In \S\ref{sec:forecast}, we provide forecasts for the detectability of the superhorizon nature of inflationary B-modes for both current and future CMB polarization experiments. 
Our conclusions are stated in \S\ref{sec:conclusion}. 

A few appendices contain additional reference materials:
 Appendix~\ref{sec:harmonic} provides details of a similar analysis in harmonic space, Appendix~\ref{sec:EffectiveNoise} describes the derivation of the effective noise in multi-frequency experiments, and Appendix~\ref{sec:Experiments} lists the instrumental specifications of the CMB experiments considered in this work.

All CMB spectra are computed with 
 CAMB~\cite{Lewis:1999bs} using the best-fit parameters of the $\Lambda$CDM model~\cite{Ade:2013zuv}: $h=0.67$, $\Omega_b h^2 = 0.022$, $\Omega_c h^2=0.12$, $\tau=0.093$, $A_s=2.2\times 10^{-9}$, and $n_s=0.96$.
The primordial tensor spectrum is taken to be scale-invariant, $n_t = 0$.

\section{The Signal}
\label{sec:signal}

We begin with a brief review of the superhorizon signature of inflationary B-modes~\cite{Baumann:2009mq}.

\subsection{Local B-modes}\label{sec:bmodes}

The polarization of the CMB is characterized by a symmetric, traceless rank-2 tensor defined in the plane perpendicular to the line-of-sight $\hat{\n}$:
\begin{equation}
P_{ij} = U\sigma_{ij}^{(1)} + Q\sigma_{ij}^{(3)}\ ,
\end{equation}
where $\sigma_{ij}^{(I)}$ denotes the Pauli matrices. Since the Stokes parameters $Q$ and $U$ transform non-trivially under rotations of the coordinates, it is more convenient to work with two invariants that can be constructed from the polarization tensor: a scalar ${\cal E} \equiv \nabla_i\nabla_j P_{ij}$ and a pseudo-scalar ${\cal B}\equiv \epsilon_{kj}\nabla_k\nabla_i P_{ij}$, corresponding to the gradient and curl parts of the polarization tensor, respectively. In the flat-sky limit, these $\E$- and $\B$-modes are related to the Stokes parameters and the ordinary $E$- and $B$-modes by~\cite{Zaldarriaga:1998rg}
\begin{align}
{\cal E}(\x) &= \nabla^2 E(\x) =(\partial_x^2 - \partial_y^2)Q(\x) + 2\partial_x\partial_y U(\x)\ ,\\
{\cal B}(\x) &= \nabla^2 B(\x) = (\partial_x^2 - \partial_y^2)U(\x) - 2\partial_x\partial_y Q(\x)\ .\label{Bdef}
\end{align}
By construction, $\E$ and $\B$ are local functions of the Stokes parameters, whereas $E$ and $B$ are defined non-locally in terms of $Q$ and $U$. Being just a linear transformation of the conventional $B$-modes, the local $\B$-modes are also a signature of tensor (and vector) modes in the initial conditions.

Any scalar field on the celestial sphere can be expanded in terms of spherical harmonics, so we write
\begin{align}
X(\hat\n) &\equiv \sum_{\ell m} a_{X,\ell m}Y_{\ell m}(\hat\n)\ ,
\end{align}
where $X=\{T, \E,\B\}$. Assuming statistical isotropy, the two-point statistics of the multipole moments are described in terms of the angular power spectrum:
\beq
\langle a_{X,\ell m\phantom{'}} \hskip -2pt a_{X,\ell' m'}^* \rangle = C_\ell^{X} \delta_{\ell \ell'} \delta_{mm'}\ ,
\eeq
where the angle brackets denote the ensemble average. 
The late-time power spectrum, $C_\ell^X$, can be related to quantum zero-point fluctuations in both the spacetime metric and the matter fields during inflation~\cite{Baumann:2009ds}. 
Given a measurement of the harmonic coefficients $a_{X,\ell m}$, we define estimators of the angular power spectra as
\beq
\widehat C_\ell^X \equiv \frac{1}{2\ell+1} \sum_m a_{X,\ell m}^{\phantom{*}}  a_{X,\ell m}^*\ .
\eeq

The power spectrum of the local $\B$-modes is related to that of the conventional $B$-modes by~\cite{Baumann:2009mq, Durrer}
\begin{equation}
C_\ell^{\B}=n^2_\ell C_\ell^B\ ,
\end{equation}
where $n_\ell \equiv \sqrt{(\ell+2)!/(\ell-2)!}$\hskip2pt. A harmonic transformation gives the corresponding correlation function in real space:
\begin{equation}\label{corr}
C^\B (\theta) = \sum_{\ell} \frac{2\ell+1}{4\pi} C^\B_\ell P_\ell(\cos\theta)\ ,
\end{equation}
where $\theta$ is the angle between pairs of line-of-sight directions $\hat{\n}_1$ and $\hat{\n}_2$, i.e.~$\cos \theta \equiv \hat{\n}_1 \cdot \hat{\n}_2$. The relation between (\ref{corr}) and the correlation function of the conventional $B$-modes is 
\begin{equation}\label{BtildeCorr}
C^{\B}(\theta) =\nabla^2 (\nabla^2+2) C^B(\theta)\ .
\end{equation}
Again, the non-local nature of the ordinary $B$-modes is manifest: (\ref{BtildeCorr}) implies that $C^{\B}$ vanishes for any $C^B$ living in the kernel of $\nabla^2(\nabla^2+2)$, even if $C^B$ is non-zero.

\subsection{Superhorizon Signal}
\label{sec:supsignal}

Having defined the local $\B$-modes, we can analyze causality constraints on their correlation functions.
The superhorizon part of the two-point correlation function is identified most directly in real space:
\begin{equation}\label{supsignal}
S^\B(\theta) \equiv H(\theta-\theta_c) \hskip 1pt C^\B(\theta)\ ,
\end{equation}
where $H$ is the Heaviside step function and $\theta_c \simeq 2.3^\circ$ is (twice) the angle subtended by the particle horizon at recombination.
The corresponding signal in harmonic space is
\begin{align}
S_\ell^\B &\,=\, 2\pi \int_{-1}^1 \d \cos \theta \, \, S^\B(\theta)\, P_{\ell}(\cos\theta) \nonumber \\[4pt]
&\,=\, \sum_{\ell'} M_{\ell\ell'} C_{\ell'}^\B \ , \label{powerspectrum}
\end{align}
where the mode-coupling matrix $M_{\ell\ell'}$ is 
\begin{equation}\label{couplingmtx}
M_{\ell\ell'} \equiv \frac{2\ell'+1}{2} \underbrace{\int_{-1}^{x_c} P_\ell(x)P_{\ell'}(x)\hskip 1pt \d x}_{\equiv\, I_{\ell\ell'}}\ ,
\end{equation}
with $x_c \equiv \cos \theta_c$.
We label the complementary subhorizon signal as $(S^\B_\ell)^\dagger \equiv C_\ell^\B - S^\B_\ell$.
The mode-coupling integrals $I_{\ell\ell'}$ in (\ref{couplingmtx}) can be calculated analytically. The off-diagonal terms are given by
\begin{align}
I_{\ell\ell'} &= \frac{ (\ell-\ell') x_c P_\ell P_{\ell'} + \ell' P_\ell P_{\ell'-1} - \ell P_{\ell-1} P_{\ell'} }{\ell(\ell+1)-\ell'(\ell'+1)}\ ,
\end{align}
where the Legendre polynomials are evaluated at $x_c$, while the diagonal terms are determined by the recursion relation
\begin{equation}\label{Idiag}
I_{\ell\ell} = \frac{2\ell-1}{2\ell+1}I_{\ell-1,\ell-1} + \frac{2\ell-1}{2\ell+1}\frac{\ell+1}{\ell}I_{\ell+1,\ell-1} - \frac{\ell-1}{\ell}I_{\ell,\ell-2}\ .
\end{equation}
We can think of the kernel~(\ref{couplingmtx}) as an operator projecting the power spectrum onto its superhorizon subspace. 
In fig.~\ref{fig:powerspectrum}, we show the superhorizon part of the power spectrum predicted by inflation. 
We see that the features of the real space correlation function above $\theta_c$ are encoded in the oscillations of the power spectrum with the frequency of the oscillations corresponding to the horizon size at recombination.

\section{The Noise}\label{sec:noise}

Next, we describe the sources of noise that we will take into account in our analysis.

\subsection{Instrumental Noise}

We represent instrumental noise by an uncorrelated Gaussian random field. 
Assuming white noise in the Stokes parameters, the noise power spectrum for $B$-modes can be expressed as~\cite{Knox:1995dq}
\begin{equation}\label{NlB}
N_\ell^B = \Delta_P^2 e^{\ell(\ell+1)/\ell_b^2}\ , 
\end{equation}
where $\Delta_P$ is the noise level of polarization sensitive detectors. 
The exponential factor in (\ref{NlB}) represents the effect of deconvolving the Gaussian beam effect from the signal, with $\ell_b \equiv\sqrt{8\ln 2}/\theta_b$ and $\theta_b$ the full width at half maximum of the beam. The noise level is determined by
\begin{equation}
\Delta_P^2 = \frac{2\text{NET}^2\Omega_\text{sky}}{N_\text{det}t_\text{obs}Y} \equiv s_P^2 f_\text{sky}\ , \label{noise2}
\end{equation}
where NET is the noise equivalent temperature of detectors, $N_\text{det}$ denotes the number of detectors, $t_\text{obs}$ is the time of observation, $Y$ characterizes the detector yield, and $\Omega_\text{sky} = 4\pi f_\text{sky}$ is the observed sky area.\footnote{The current generation of experiments achieves $ \text{NET} = 350 \mu{\rm K}\sqrt{s}$ and $Y=0.25$, with $N_{\rm det} \sim {\cal O}(10^3)$~\cite{Wu:2014hta}. 
 In \cite{Abazajian:2013oma} ground-based experiments have been classified by the number of detectors as Stage-II, Stage-III and Stage-IV for $N_{\rm det} \sim 10^3$, $10^4$ and $10^5$, respectively.}
We will find it useful to consider the effective sensitivity of a full-sky experiment, $s_P$, and rescale it by the observed sky fraction, $f_{\rm sky}$. In an experiment with multiple frequency channels, a heuristic measure of the effective noise level  is
\begin{equation}
\Delta_{P,\text{eff}}^2 = \left[\sum_{i} \frac{1}{\Delta_{P,i}^2}\right]^{-1}\ ,
\end{equation}
where $\Delta_{P,i}$ denotes the noise level of channel $i$. The instrumental specifications and the effective noise levels for a selection of current and upcoming B-mode experiments are listed in Table \ref{tab:specs} and in more detail in Appendix~\ref{sec:Experiments}. 
\begin{table}[t!]
\centering
\begin{tabular}{ l c r c c }
\hline
& $\theta_b\,[']$ & $f_\text{sky}$ [\%] & $\Delta_{P,\text{eff}}\,[\mu\text{K}']$ & $s_{P,\text{eff}}\,[\mu\text{K}']$ \\ [0.5ex] 
\hline 
BICEP2 & 29  & 2.4\hphantom{11} & \hphantom{1}5.2 & 33.6 \\
Keck Array & 29 & 2.4\hphantom{11} & \hphantom{1}2.2 & 14.2\\
PolarBeaR-2 & \hphantom{1}4 & 20\hphantom{11} & 10.7 & 23.9\\
Simons Array & \hphantom{1}3 &  20\hphantom{11} & \hphantom{1}6.3 & 14.1\\
SPTPol & \hphantom{1}1 & 6\hphantom{11} & \hphantom{1}4.4 & 17.8\\
LiteBIRD & 16 & 70\hphantom{11} & \hphantom{1}1.8 &  \hphantom{1}2.2 \\
COrE & \hphantom{1}1 &  70\hphantom{11} & \hphantom{1}1.8 & \hphantom{1}2.2 \\[1ex] 
\hline\hline
\end{tabular} 
\caption{Instrumental specifications for current and upcoming CMB polarization experiments~\cite{Ade:2014gua, Bock, Kermish:2012eh, Bouchet:2011ck, Austermann:2012ga, doi:10.1117/12.926158, Matsumura:2013aja}.}
\label{tab:specs}
\end{table} 

As can be seen from (\ref{Bdef}), a measurement of $\B$-modes effectively involves taking derivatives of both the signal and the noise. This has the benefit that the observables become local quantities, but at the same time the noise spectrum for $\B$-modes acquires a factor of $n_\ell^2 \sim \ell^4$ relative to the noise for a $B$-mode measurement. White noise spectra for the Stokes parameters then translate into a blue spectrum for $\B$-modes, $N_\ell^\B \sim \ell^4$, implying a large contribution from small-scale noise.
 Because of the drastic difference in the properties of the noise, it is important to analyze the detectability of $\B$-modes separately, adopting a different strategy from measurements of $B$-modes if necessary.
 
In order to compensate for the blue noise spectrum, we will apply a low-pass filtering to both the signal and the noise:\hskip 2pt\footnote{Calligraphic font will from now on denote filtered quantities.} 
\begin{equation}
{\cal C}_\ell^\B \equiv f_\ell C_\ell^\B \ ,\quad  
 {\cal N}_\ell^\B \equiv f_\ell N_\ell^\B\ , \label{smoothing}
\end{equation}
where $f_\ell$ denotes a filtering function. 
In real space, the procedure~(\ref{smoothing}) corresponds to a convolution with a certain window function $f(\theta,\theta') = \sum_\ell \frac{2\ell+1}{2}f_\ell \hskip 1pt P_\ell(\cos\theta)P_\ell(\cos\theta')$. 
Depending on the experimental strategy, different window functions may be more suitable. For our purposes, there are several conditions that the filtering function $f_\ell$ needs to satisfy: (i) it needs to be sufficiently smooth to avoid the Gibbs phenomenon, (ii) it should decay early enough to suppress the small-scale noise efficiently, and (iii)~it should retain the shape of the power spectrum up to $\ell\sim 100$ in order not to cause any distortion of the superhorizon features. A simple choice which satisfies the above requirements is a Gaussian filtering function:
\begin{equation}
f_\ell = e^{-\ell(\ell+1)/\ell_s^2}\ , \label{equ:Fell}
\end{equation}
where $\ell_s$ defines the smoothing scale. 
To satisfy the second and third conditions, we choose $100<\ell_s<\ell_b$, in which case the first condition is automatically satisfied.

\subsection{Leakage}\label{sec:leakage}

The filtering of the $\B$-mode spectrum is a necessary evil.
An inevitable consequence of the filtering process is a transfer of part of the subhorizon signal to superhorizon scales (and vice versa). For lack of a better term, we will call this contamination {\it leakage}. Since the spurious modes due to leakage can confuse the detection of the true superhorizon signal, it will be important to treat them carefully in our analysis. 
 
\begin{figure}[t!]
\centering
\includegraphics[width=0.47\textwidth]{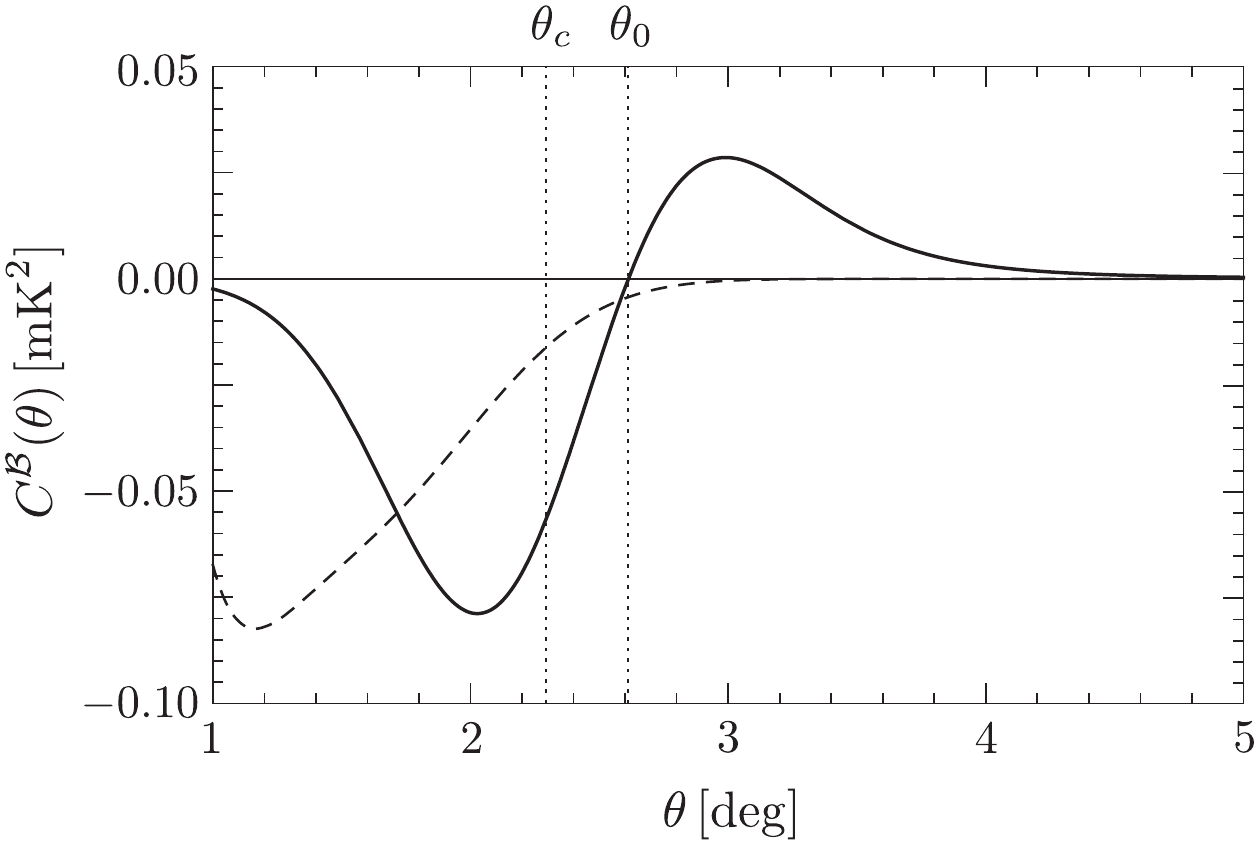}
\caption{Local $\B$-mode correlation function for $r=0.13$ using the Gaussian filter (\ref{equ:Fell}) with $\ell_s=200$. The solid and dashed lines correspond to the superhorizon and subhorizon signals, respectively. }
\label{fig:CBfiltered}
\end{figure}

In fig.~\ref{fig:CBfiltered}, we show the filtered subhorizon and superhorizon $\B$-mode correlation functions. As we can see, there is a non-negligible amount of leakage around $\theta\sim 2^\circ$. On the other hand, the positive peak of the superhorizon signal at $\theta\sim 3^\circ$ is relatively clean and still serves as an unambiguous test of the inflationary superhorizon spectrum. 
When we want to make sure that we don't suffer from a large amount of leakage, we therefore focus on correlations with $\theta \gtrsim \theta_0 \equiv 2.6^\circ$.
Moreover, at fixed $\theta$, the leakage can be reduced by working with larger values of~$\ell_s$.
However, making $\ell_s$ too large will reduce the signal-to-noise of the signal we wish to measure.
In \S\ref{sec:forecast}, we will discuss the optimal balance between minimal leakage and maximal signal-to-noise.

%

\subsection{Foregrounds}

Our ability to detect the primordial B-mode signal depends crucially on how well we can separate the signal from foreground contamination. The two major sources of foregrounds in the microwave range are polarized emissions from synchrotron and thermal dust. 
Their distinct frequency dependences, in principle, allow them to be distinguished from the primary CMB signal.

\subsubsection{Synchrotron}

Synchrotron radiation arises from the acceleration 
of relativistic cosmic-ray electrons in the magnetic field of the Galaxy.
This is the dominant contribution to the polarized foreground emission below 70 GHz. 

\vskip 4pt
If the electrons have a power law distribution of energies, $N(E) \propto E^{-p}$, then the antenna temperature\footnote{Antenna temperature units are defined in reference to the Rayleigh-Jeans law, whereas thermodynamic temperature units are defined as the blackbody temperature obeying Planck's law. We calibrate quantities in thermodynamic temperature units, so that the primary CMB spectrum is frequency-independent.} of the signal is predicted to have a power-law dependence on frequency, $T(\nu) \propto \nu^{\beta_s}$, with $\beta_s = -\frac{1}{2}(p+3)$. 
This simple ansatz for the frequency spectrum fits observations rather well with $\beta_s\simeq -2.9 $~\cite{Ade:2014zja}.
The variation of the spectral index across the sky is of order 10\%.
The angular spectrum of the synchrotron emission is found to obey an approximate power law, $F_\ell^{s,B} \propto \ell^{\alpha_s}$, with  $\alpha_s \simeq -2.6$~\cite{Page:2006hz}.

Combining the above facts, we are led to the following ansatz
for the synchrotron $B$-mode power spectrum in thermodynamic temperature units~\cite{Baumann:2008aq}:
\begin{equation}
F_\ell^{s,B}(\nu) =  A_s \left(\frac{\ell}{\ell_0}\right)^{\alpha_s} h^s(\nu,\nu_0)\ , \label{Fs}
\end{equation}
where $A_s$ is the amplitude of synchrotron emission defined at a reference frequency $\nu_0$ and a reference scale~$\ell_0$. The function
\begin{equation}
h^s(\nu,\nu') \equiv \left(\frac{\nu}{\nu'}\right)^{2\beta_s}  \left(\frac{f(\nu)}{f(\nu')}\right)^2
\end{equation}
encapsulates the spectral dependence, where the factor $f(\nu)$ accounts for the conversion from antenna temperature to thermodynamic temperature~\cite{Tegmark:1999ke}:
\begin{equation}
f(\nu) \equiv \frac{(e^x - 1)^2}{x^2 e^x}\ , \quad x\equiv \frac{h\nu}{kT_{\rm cmb}} \approx 
\frac{\nu}{56.8 \,{\rm GHz}}\ ,
\end{equation}
where  $T_{\rm cmb}=2.725\, {\rm K}$ is the CMB blackbody temperature~\cite{Fixsen:2009ug}.

\subsubsection{Thermal Dust}

Thermal emission from interstellar dust grains aligned with the Galactic magnetic field
produces the dominant polarized foreground above 70 GHz. 

\vskip 4pt
The frequency dependence of the dust intensity takes the form of a modified blackbody, $I_\nu \propto \nu^{\beta_d} B_\nu(T_d) $, where
the Planck spectrum $B_\nu(T_d)$ is determined by the observed dust temperature, $T_d\simeq 19.7\, {\rm K}$~\cite{Abergel:2013fza}.  The mean spectral index is found to be $\beta_d \simeq 1.5$ at microwave frequencies~\cite{Ade:2014zja}, with a variation of about 1\% across the sky (much less than the variation of the synchrotron spectral index). 
The angular spectrum again satisfies a power law,
$F_\ell^{d,B} \propto \ell^{\alpha_d}$, with $\alpha_d \simeq -2.3$~\cite{Aumont}.
The dust $B$-mode power spectrum can therefore be modelled as~\cite{Baumann:2008aq}
\begin{equation}
F_\ell^{d,B}(\nu) = A_d  \left(\frac{\ell}{\ell_0}\right)^{\alpha_d}  h^d(\nu,\nu_0) \ , \label{Fd}
\end{equation}
where $A_d$ is the amplitude of the polarized dust emission defined at a reference frequency $\nu_0$ and a reference scale~$\ell_0$. The spectral function for dust is\hskip1pt\footnote{Eq.~(\ref{hd}) corrects a typo in~\cite{Tucci:2004zy,Verde:2005ff}.}
\begin{equation}\label{hd}
h^d(\nu,\nu')\equiv \left(\frac{\nu}{\nu'}\right)^{2\beta_d} 
\left(\frac{B_\nu(T_d)}{B_{\nu'}(T_d)} \frac{g(\nu)}{g(\nu')}\right)^2\ ,
\end{equation}
where $g(\nu)$ is the conversion factor from intensity to thermodynamic temperature units~\cite{Tegmark:1999ke},
\begin{equation}
g(\nu) \equiv \frac{f(\nu)}{\nu^2} \ .
\end{equation}

The amplitude in (\ref{Fd}) can be written as $A_d = p^2 I_d$, where $I_d$ is the unpolarized dust intensity and $p$ is the polarization fraction. 
Both $p$ and $I_d$ can vary significantly across the sky.\footnote{The latest Planck measurements suggest that the mean polarization fractions over most parts of the sky (including highly polarized regions) fall in the range of  3 to 14\%~\cite{Ade:2014gna}. The dust intensity is constrained by the Finkbeiner-Davis-Schlegel (FDS) dust map~\cite{Finkbeiner:1999aq}.}
The precise amount of foreground contamination, therefore, depends on the region of the sky under consideration. 
In $\S$\ref{sec:forecast}, we will consider a few different choices for the amplitude of the dust polarization, and allow for the relatively large uncertainties that still exist.

\subsubsection{Foreground Residuals}

Multi-frequency observations allow some degree of foreground cleaning based on the distinct frequency dependence of the foregrounds. 
Detailed algorithms for foreground cleaning are discussed in \cite{Dunkley:2008am}. 
Following~\cite{Verde:2005ff, Baumann:2008aq}, we will assume that the foregrounds can be subtracted by the template cleaning method (e.g.~\cite{Eriksen:2005dr,Stompor:2008sf}), and simply parameterize the foreground residuals by rescaling the foreground amplitudes by two scale-independent factors, $\epsilon_x \in [0,1]$, with $x=\{s,d\}$ denoting synchrotron and dust, respectively. 
We propagate the noise of the template map into the foreground residuals.
After cleaning, the residual foreground spectrum, then, is~\cite{Tucci:2004zy,Verde:2005ff}
\beq\label{RlB}
R_\ell^B \equiv \sum_x \left[\epsilon_x F^{x,B}_\ell + {\sf N}_\ell^{x,B} h^x(\nu,\nu_{\rm ref}^x)\right]\ ,
\eeq
where $\nu^x_{\rm ref}$ is the reference frequency used as the template and ${\sf N}_\ell^{x,B}$ is the noise level of the template map for $x$.

We treat the foreground residuals as additional sources of uncorrelated noise (see Appendix~\ref{sec:EffectiveNoise} for a discussion).  For an experiment with multiple frequency channels, we seek to find a linear combination of the maps with weightings chosen in such a way to minimize the variance of the power spectrum~\cite{Tegmark:1999ke}. 
In Appendix~\ref{sec:EffectiveNoise}, we derive the optimal weighting scheme and show that the effective noise of the combined map is~\cite{Baumann:2008aq}
\begin{equation}
\label{Neff}
N_{{\rm eff},\ell}^B = \left[ \sum_{i} \frac{1}{N_{i,\ell}^B + R_{i,\ell}^B }\right]^{-1}\ ,
\end{equation}
where the subscript $i$ denotes the value at frequency $\nu_{i}$.
Appendix~\ref{sec:EffectiveNoise} also explains that any correlations between the foreground residuals at different frequencies tend to reduce the effective noise level, so working with (\ref{Neff}) is a conservative choice.

\subsection{Lensing}

Even in the absence of primordial B-modes, a curl component of CMB polarization is generated by the lensing of primordial E-modes~\cite{Zaldarriaga:1998ar, Hu:2000ee}.
This effect has to be considered an additional source of noise for the signal we are trying to measure.

\vskip 4pt
On large angular scales, the lensing $B$-modes act like white noise with an effective amplitude of $4.4\, \mu{\rm K}'$. In the low-noise regime ($\lesssim 5\, \mu\text{K}'$), the lensing effect provides a significant limitation to a measurement of the primordial signal, especially for low values of $r$. Since lensing does not induce any spectral distortions to the primary CMB, multi-frequency observations do not help to distinguish between these two signals. However, several methods have been proposed to reduce the lensing noise statistically~\cite{Okamoto:2003zw, Hirata:2003ka} (see \cite{Smith:2008an} for a comprehensive discussion). The most promising delensing procedure involves reconstructing the lensing potential from measurements of small-scale CMB polarization, which is subsequently used to remove the lensing contribution to the large-scale B-mode signal. This requires CMB experiments with high sensitivity and resolution (small beam size). Details of this approach to delensing can be found e.g.~in~\cite{Smith:2010gu}. 

In the absence of sky cuts, foregrounds, and instrumental systematics, a detection of the primordial tensor amplitude down to $r\sim10^{-6}$, in principle, is achievable~\cite{Seljak:2003pn}. Nevertheless, experimental limitations and the presence of foregrounds practically limit an accurate quantification of the residual lensing, resulting in a possible bias in the estimator of the lensing potential. To avoid these practical uncertainties, we assume that the lensing estimator is unbiased, or that any significant biases are known and can be eliminated. Thus, the residual lensing contributes only to the variance, and does not bias the signal. The issue of potential lensing bias is the subject of many investigations in the literature, e.g.~\cite{Shimon:2007au,Hanson:2009dr,Hanson:2010gu,Hanson:2010rp,Namikawa:2012pe,BenoitLevy:2013bc,vanEngelen:2013rla}, but is beyond the scope of the present work. 

We consider delensing in a heuristic way by multiplying the amplitude of the lensing $B$-modes $L_\ell^B$ by a scale-independent delensing fraction,
\beq
L_\ell^B \ \to\ \epsilon_L L_\ell^B\ , \label{equ:lensing}
\eeq
where $\epsilon_L \in [0,1]$, and treat it as an additional noise. On large scales, both the residual spectrum and the original spectrum are approximately white noise (see e.g.~fig.~1 in \cite{Boyle:2014kba}). Therefore, the ansatz (\ref{equ:lensing}) is a sufficiently good approximation to more sophisticated expressions for the lensing residuals found in Appendix~A of \cite{Smith:2010gu}. The residual lensing power spectrum is then incorporated into the effective noise as 
\begin{equation}
N_{{\rm eff},\ell}^B = \left[ \sum_{i} \frac{1}{{N}^{B}_{i,\ell} + R_{i,\ell}^B}\right]^{-1} + \epsilon_L L_{\ell}^B \ .
\end{equation}
Further justification for this formula is given in Appendix~\ref{sec:EffectiveNoise}.

\section{Methodology}
\label{sec:method}

We now describe our method for quantifying the detectability of the superhorizon $\B$-mode signal.  We first construct an estimator of the signal and then use it to define the signal-to-noise ratio of the measurement.  We will explain that leakage introduces a bias in the estimator and describe a simple debiasing procedure.
The methodology in this section and the next will be formulated mostly  in real space, but see Appendix~\ref{sec:harmonic} for an equivalent treatment in harmonic space.

\subsection{Superhorizon Estimator}

We would like to define an estimator of the superhorizon signal~(\ref{supsignal}), 
given an estimator $\widehat{\cal C}_\ell$ for the total $\B$-mode power spectrum after filtering, $\langle \widehat{\cal C}_\ell\rangle = {\cal C}_\ell^\B \equiv f_\ell C_\ell^\B$. The associated covariance matrix is\hskip 1pt \footnote{Lensing induces a non-Gaussian contribution to the covariance matrix whose explicit expression can be found in~\cite{Smith:2004up}. We have checked that the degradation caused by the non-Gaussian lensing covariance is much smaller than the systematic uncertainties due to the leakage.}  
\begin{equation} \label{equ:CovMat}
\mathscr{C}[\widehat{\cal C}_\ell,\widehat{\cal C}_{\ell'}] = \frac{2}{(2\ell+1) f_\text{sky}}   \left({\cal C}^\B_\ell + {\cal N}^\B_{{\rm eff},\ell}\right)^2\, \delta_{\ell\ell'}\ .
\end{equation}
Selecting the total signal in the angular interval $\Theta \equiv [\theta_{\rm min}, \theta_{\rm max}]$, with $\theta_{\rm min} \ge \theta_c$, defines
an estimator of the superhorizon signal\hskip 1pt\footnote{In Appendix~\ref{sec:harmonic}, we define the harmonic space equivalent of the estimator (\ref{realestimator}).}
\begin{equation}\label{realestimator}
\widehat{\mathcal{S}}(\theta; \theta_{\rm min}) \,\equiv\, \sum_\ell \frac{2\ell+1}{4\pi}\hskip 1pt \widehat{\cal C}_\ell \hskip 1pt P_\ell(\cos\theta)\hskip 1pt \Pi(\theta)\ ,
\end{equation}
where
\begin{equation}
\Pi(\theta) \equiv \left\{ \begin{array}{ll} 1 &\quad \theta \in [\theta_{\rm min}, \theta_{\rm max}] \\[6pt] 0 &\quad  \text{otherwise} \end{array} \right. \ .
\end{equation}
For now, we will keep $\theta_{\rm min}$ general. The precise definition of $\theta_{\rm max}$ is not important, but will be limited by the maximum angular extent of a partial sky observation.
The covariance of the estimator is given by
\begin{align}\label{covreal}
\mathscr{C}[\widehat{\mathcal{S}}(\theta),\widehat{\mathcal{S}}(\theta')] &= \sum_{\ell\ell'} \frac{2\ell+1}{4\pi}\frac{2\ell'+1}{4\pi} \mathscr{C}[\widehat{\cal C}_\ell,\widehat{\cal C}_{\ell'}] \nonumber\\
&\quad \times  P_\ell(\cos\theta)P_{\ell'}(\cos\theta')\hskip 1pt\Pi(\theta)\Pi(\theta')\ .
\end{align}
We emphasize that the estimator (\ref{realestimator}) is {\it biased}, since the total signal contains spurious contributions from the filtered subhorizon modes (see~\S\ref{sec:leakage}).
In \S\ref{sec:bias}, we will quantify this bias and define a debiased version of the estimator.

\subsection{Signal-to-Noise}
\label{sec:snr}

To define the signal-to-noise of the measurement, we discretize (\ref{realestimator}) and (\ref{covreal}), and split the signal into $N$ uniformly spaced angular bins $\Theta_{b} \equiv \{\theta_{(b)} \pm \frac{1}{2}\Delta \theta \}$, for $b=1,\ldots, N$. A natural sampling interval is $\Delta\theta \simeq 180^\circ/\ell_\star$, where $\ell_\star$ is the multipole moment at which the covariance matrix~(\ref{covreal}) converges.\footnote{The convergence of (\ref{covreal}) at $\ell_\star$ means that we effectively take into account $\ell_\star$ independent modes of ${\cal C}_\ell^\B$, in which case the rank of the matrix (\ref{equ:CovMat}) is $\ell_\star$. Since the transformation from harmonic space to real space is linear, the rank of the corresponding covariance matrix in real space is also $\ell_\star$. By restricting to a proper subinterval, $\Theta \equiv [\theta_{\rm min}, \theta_{\rm max}]$, we effectively reduce the rank by a factor of $\sim 180^\circ/(\theta_{\rm max} - \theta_{\rm min})$. Thus, a natural sampling interval is $\Delta\theta = 180^\circ/\ell_\star$. (In practice, the optimal $\Delta\theta$ is slightly larger, since the signal decays before it reaches $\ell_\star$ and we also include the non-Gaussian part of the covariance.) Errors at different angular separations are strongly correlated within the interval $\Theta$, and oversampling will result in an ill-behaved covariance matrix.}
The average signal assigned to each bin is
\beq
\widehat{\mathcal{S}}_{b} \equiv \frac{1}{Z_{b}}\int\limits_{\Theta_{b}} \d\theta\sin\theta\,\, \widehat{\mathcal{S}}(\theta)\ ,
\eeq
where $Z_{b} \equiv \int_{\Theta_{b}} \d \theta\, \sin \theta$ is a normalization factor.
The binned covariance matrix is given by
\begin{equation}\label{covmtxreal}
\mathscr{C}_{bb'} \equiv \frac{1}{Z_{b} Z_{b'}} \int\limits_{\Theta_{b}}\int\limits_{\Theta_{b'}} \d \theta \d \theta' \sin\theta\sin\theta'\, \mathscr{C}[\widehat{\mathcal{S}}(\theta),\widehat{\mathcal{S}}(\theta')]\ ,
\end{equation} 
and the signal-to-noise ratio is defined as
\begin{equation}\label{SNR}
\left({\rm S/N}\right)^2 = \sum_{bb'} \widehat{\mathcal{S}}_{b} \hskip 1pt\mathscr{C}^{-1}_{bb'} \widehat{\mathcal{S}}_{b'}\ ,
\end{equation}
where $\mathscr{C}^{-1}_{bb'}$ is the inverse of (\ref{covmtxreal}).
In \S\ref{sec:forecast}, we will evaluate (\ref{SNR}) for various experimental configurations.

\subsection{Leakage and Debiasing}
\label{sec:bias}

Since the total signal ${\cal S}(\theta)$ contains spurious modes from the leakage of the filtered subhorizon modes, 
$\widehat{\cal S}(\theta)$ is a biased estimator of the true superhorizon signal~(\ref{supsignal}).\footnote{Another type of bias arises from the $E$-$B$ mixing in partial sky observations. This bias is well-understood and can be treated by substituting the pseudo-$C_\ell$ estimators considered in~\cite{Smith:2005gi} for $\widehat{C}_\ell^\B$.}
We will quantify this bias by comparing the signal-to-noise of the expected total signal with that of the spurious subhorizon modes. 

\vskip 4pt
Let us write the estimator~(\ref{realestimator}) as
$\widehat{\cal S}(\theta) = \widetilde{\cal S}(\theta) + {\cal S}^\dagger(\theta)$, where $ \widetilde{\cal S}(\theta)$ denotes the unbiased estimator (i.e.~the estimator of the pure superhorizon component) and ${\cal S}^\dagger(\theta)$ is the subhorizon signal.
The total signal-to-noise (\ref{SNR}) can then be written as
\begin{align}
\left({\rm S/N}\right)^2 &= \sum_{bb'} \left( \widetilde{\cal S}_{b} \hskip 1pt\mathscr{C}^{-1}_{bb'} \widetilde{\cal S}_{b'} + 2\hskip 1pt\widetilde{\cal S}_{b} \hskip 1pt\mathscr{C}^{-1}_{bb'} {\cal S}^\dagger_{b'}  + {\cal S}^\dagger_{b} \hskip 1pt\mathscr{C}^{-1}_{bb'} {\cal S}^\dagger_{b'} \right) \nonumber \\[2pt]
&\equiv ({\rm S}/{\rm N})^2_+ + ({\rm S}/{\rm N})^2_\times + ({\rm S}/{\rm N})^2_-\ ,
\end{align}
where $({\rm S}/{\rm N})_+$ and $({\rm S}/{\rm N})_-$ denote the parts coming from the true superhorizon modes and the subhorizon leakage, respectively, while  $({\rm S}/{\rm N})_\times$ stands for their cross-correlation. We will use 
\begin{equation}\label{LF}
\delta \equiv \frac{({\rm S}/{\rm N})_-}{{\rm S/N}} 
\end{equation}
as a diagnostic tool for quantifying the amount of leakage and, hence, the bias in the estimator~(\ref{realestimator}). 
For small values of $\delta$, we know that the expected signal is dominated by the true superhorizon modes.  
We will consider optimizing the analysis (e.g.~by adjusting $\ell_s$ and $\theta_{\rm min}$), so that we get the maximum signal-to-noise while keeping the leakage fraction (\ref{LF}) small. 
We typically take an acceptable leakage fraction to be~$\delta \le 0.1$.

Alternatively, we can correct for the bias of the estimator~(\ref{realestimator}) through a simple debiasing procedure. 
Subtracting the expected ensemble average of the spurious subhorizon mode from the estimator~(\ref{realestimator}) leads to an unbiased estimator of the pure superhorizon signal:
\begin{equation}
\label{debiased}
\widetilde{\cal S}(\theta) \equiv \widehat{\cal S}(\theta) - {\cal S}^\dagger(\theta)\ .
\end{equation}
In this case, we can treat the subhorizon signal as an extra source of noise. 
Applying this debiasing procedure, we may improve the signal-to-noise by allowing a smaller smoothing scale $\ell_s$ and/or a larger angular interval $\Theta$.

\section{Signal-to-Noise Forecasts}
\label{sec:forecast}

Finally, we are ready to investigate the detectability of the superhorizon $\B$-mode signal for current and future experiments.  
The signal-to-noise will, of course, depend on the tensor-to-scalar ratio of the primordial fluctuations. We will consider both a fiducial value of $r=0.13$ (which corresponds to the amplitude suggested by BICEP2~\cite{Ade:2014xna} and is also the canonical value of $m^2 \phi^2$ chaotic inflation~\cite{Linde:1983gd}), as well as the wider range $r=[0.001,0.2]$. 

\subsection{Preliminaries} 

We will use the estimators (\ref{realestimator}) and (\ref{debiased}) defined on the interval $\Theta = [\theta_{\rm min}, \theta_{\rm max}]$, with $\Delta\theta=0.30^\circ$. 
For simplicity, we will fix $\theta_{\rm max} = 6.0^\circ$ throughout. 
For $\theta_{\rm min}$, we will consider two different choices: 
\begin{enumerate} 
\item[(I)] For the biased estimator~(\ref{realestimator}), we compute the signal-to-noise on an interval with $\theta_{\rm min} = 2.6^\circ$, where the leakage from subhorizon modes is guaranteed to be small and constrained by causality.  
\item[(II)] For the debiased estimator~(\ref{debiased}), we compute the signal-to-noise on an extended interval with $\theta_{\rm min} = 1.0^\circ$, which is where the filtered pure superhorizon signal\footnote{Below we will show that the Gaussian (\ref{equ:Fell}) with $\ell_s =200$ is a conservative filter function. We will take this as our fiducial choice of filtering, but also investigate the possibility of optimizing the smoothing scheme in particular examples. } starts to become appreciable (c.f.~fig.~\ref{fig:CBfiltered}). 
\end{enumerate} 
The estimator (I) is clearly more conservative, but also rejects a significant fraction of the inflationary superhorizon signal. The estimator (II), on the other hand, includes all superhorizon modes, but is less immune to spurious subhorizon contamination due to leakage. Although the known bias due to the inflationary subhorizon modes has been corrected for in the estimator~(\ref{debiased}), 
a signal on the interval $[1^\circ, 2^\circ]$ from non-inflationary sources is strictly speaking not forbidden by causality. To perform a true causality test of inflationary tensor modes, we therefore aim to detect the signal with the estimator (I). 
Nevertheless, we will also show results for the estimator~(II) which quantifies the signal-to-noise of the total superhorizon signal from inflation. In that case, the caveat that we just stated should be kept in mind. 

\vskip 4pt
We will consider two sets of foreground models: 
\begin{itemize} 
\item Ground-based experiments (\S\ref{sec:ground}) can target small, but exceptionally clean, patches of the sky, and lower estimates for the foreground amplitudes are therefore appropriate. 
\item Space-based all-sky experiments (\S\ref{sec:space}) can't use the cleanest patches only, so we will use higher foreground levels in those cases. 
\end{itemize} 
Our precise choices for the foreground amplitudes will depend on the experiment under consideration and will be presented in the following sections.

\subsection{Ground-Based Experiments}
\label{sec:ground}

We first consider the capabilities of ground-based experiments, as illustrated by a few representative examples.

\subsubsection{Keck Array} 

The BICEP2 experiment has recently been upgraded to the Keck Array~\cite{Ogburn:2012ma}.
The Keck Array, unlike BICEP2, has multiple frequency channels, and the combination of its 95, 150, and 220 GHz detectors yields an effective noise of $\Delta_{P,{\rm eff}}=2.2\, \mu{\rm K}'$ ($s_{P,{\rm eff}}=14.2\, \mu\text{K}'$).
The 95 and 150 GHz channels are already in operation, and the 220~GHz channel will be added soon. In the near future, the BICEP3 experiment~\cite{Ahmed:2014ixy} will start to observe the same part of the sky, with higher sensitivity at 95~GHz.
In combination with the Keck Array, the effective noise will then reduce to $\Delta_{P,{\rm eff}}=1.4\, \mu{\rm K}'$ ($s_{P,{\rm eff}}=9\, \mu\text{K}'$).  In the following, we will refer to this combination of the Keck Array and BICEP3 simply as the `Keck Array'.

Like for BICEP2, observations are made in the ``Southern Hole'' ($f_{\rm sky} = 0.024$), a region where both galactic and extragalactic foreground emissions are expected to be very low.   For the foreground amplitudes in the Southern Hole, we will use the estimates given in~\cite{Flauger:2014qra}: 
\begin{align}
A_s &\ =\ \xi_s \times (1.5 \times 10^{-7}\, \mu{\rm K}^2) \ , \label{As}\\
A_d &\ =\ \xi_d \times (1.8 \times 10^{-6}\, \mu{\rm K}^2)\ , \label{Ad}
\end{align}
where these amplitudes are measured at $\nu_0=100\, {\rm GHz}$ and $\ell_0=100$. 
The parameters $\xi_s$ and $\xi_d$ allow for our uncertainties concerning the synchrotron and dust amplitudes in the Southern Hole. We will use $\xi_s = [0.67,1.33]$ and $\xi_d = [0.33,1.67]$ which corresponds to the 1$\sigma$ uncertainties in~\cite{Flauger:2014qra}.

Using the 220 GHz map of the Keck Array as a template, internal foreground removal of polarized dust emission at lower frequencies will be possible to some extent. This requires the spectral index of the dust signal to be well-constrained, which will be the case if external information from Planck is folded in. Our uncertainty in the level of foreground residuals that can ultimately be achieved will be characterized by the parameters $\epsilon_s$ and $\epsilon_d$ in (\ref{RlB}).

The large beam size of the Keck Array ($\theta_b \sim 30'$) means that internal delensing will not be possible; yet a joint analysis with a higher resolution experiment observing the same part of the sky may allow some modest amount of delensing. 
SPTPol~\cite{Hanson:2013hsb} is indeed also observing in the Southern Hole, but its current sensitivity isn't at a level that would make delensing a realistic possibility. In the following, we will therefore assume Keck Array observations without any delensing as the default, i.e.~$\epsilon_L = 1$ in~(\ref{equ:lensing}), but also give results invoking a small amount of delensing, $\epsilon_L =\{0.5, 0.3\}$, as might become possible with an upgrade of SPTPol.

\subsubsection{Simons Array}

The Simons Array~\cite{Lee} is a planned successor of the PolarBeaR experiment~\cite{Kermish:2012eh, doi:10.1117/12.926158}.
Located in the Atacama desert in Chile, it will provide high-resolution observations of a relatively large fraction of the sky ($f_{\rm sky} =0.2$).
The frequency bands of the Simons Array are the same as those of the Keck Array: 95, 150, 220~GHz. The effective noise level is $\Delta_{P,{\rm eff}}=6.3\, \mu{\rm K}'$ ($s_{P,{\rm eff}}=14.1\, \mu{\rm K}'$). 

In the absence of detailed information about the polarized emission in the region observed by the Simons Array, we will use the same foreground levels (\ref{As}) and (\ref{Ad}) as for the Keck Array, with the same, relatively large, uncertainties.
Its small beam size ($\theta_b=2.7'$ at 220 GHz) allows the Simons Array to serve as a useful probe to the gravitational lensing of the CMB on small angular scales, and internal delensing will be possible to some degree. We will thus show results for $\epsilon_L = \{0.5, 0.3\}$.

\subsubsection{Results}

\begin{figure}[h]
\centering
\includegraphics[width=0.45\textwidth]{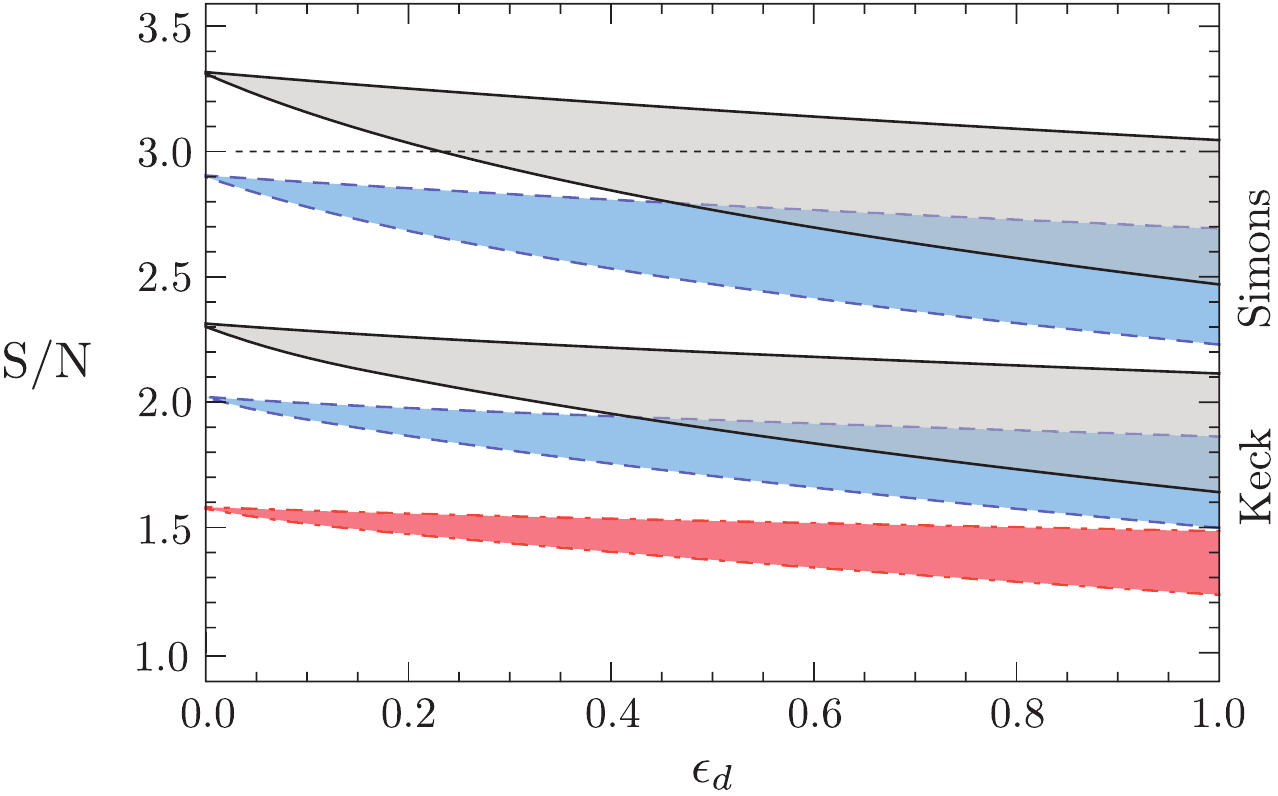}
\vskip -4pt
\caption{Signal-to-noise on the interval $[2.6^\circ,6.0^\circ]$ for $r=0.13$ as a function of $\epsilon_d$. The plot shows experiments with Keck Array (bottom) and Simons Array (top) specifications for three different delensing fractions: $\epsilon_L=1.0$ (red, dot-dashed),  $\epsilon_L=0.5$ (dashed, blue), and $\epsilon_L=0.3$ (solid, black). The bands correspond to the uncertainty in the foreground amplitudes, $\xi_s=[0.67,1.33]$ and $\xi_d =[0.33,1.67]$.}
\label{fig:SNR}
\end{figure}

In fig.~\ref{fig:SNR}, we present results for the signal-to-noise achievable by the Keck Array and the Simons Array for the fiducial value $r=0.13$ as a function of the level of foreground cleaning $\epsilon_d$. Shown are various levels of the delensing fraction $\epsilon_L = \{1,0.5,0.3\}$.
We see that a $3\sigma$ detection will marginally be possible with the Simons Array if both delensing and foreground cleaning can be achieved to a relatively high standard.
On the other hand, a detection with the Keck Array does not look feasible.

\begin{figure}[h]
\centering
\includegraphics[width=0.45\textwidth]{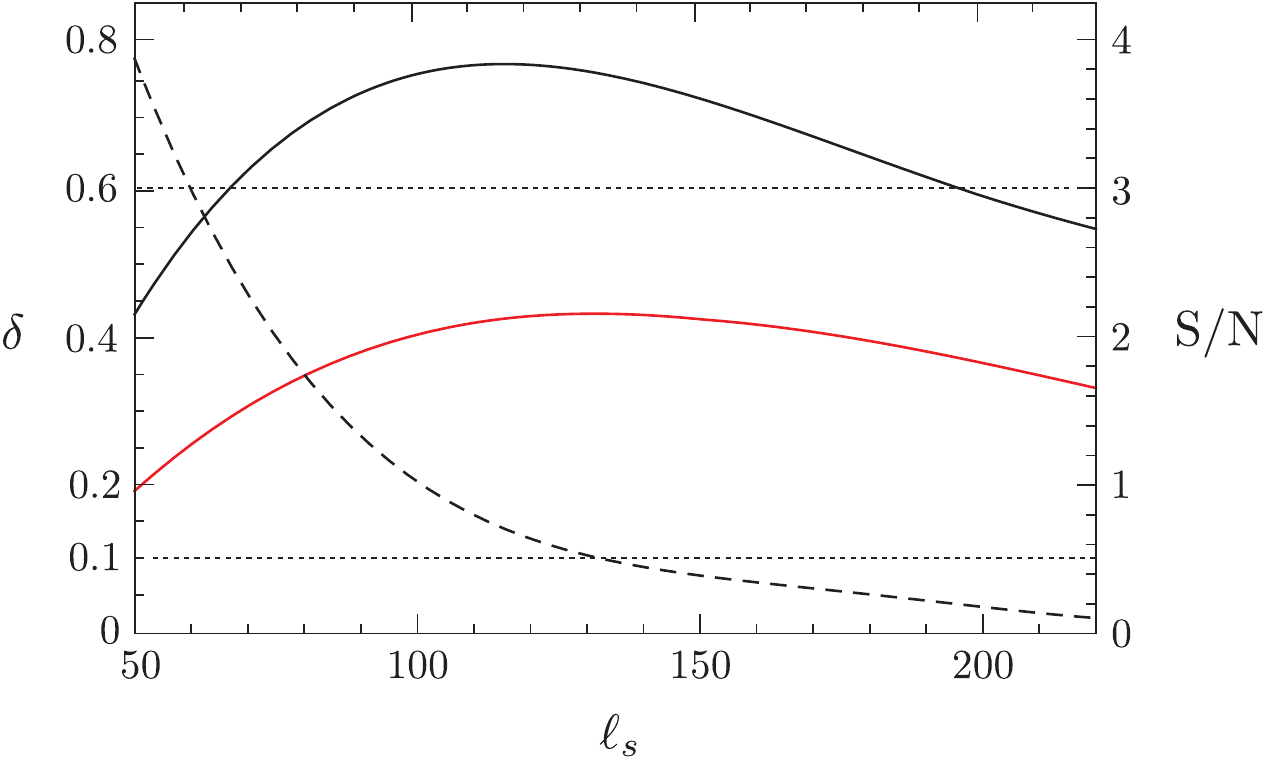}
\vskip -4pt
\caption{Signal-to-noise (solid) and leakage fraction (dashed) for $r=0.13$ as a function of $\ell_s$ for experiments with Keck Array (red) and Simons Array (black) specifications. 
Only a single curve is shown for $\delta$ because the curves for the Keck Array and the Simons Array are almost identical.
The plot assumes $\epsilon_L=\{\epsilon_s,\epsilon_d\}=0.5$. Decreasing the smoothing scale from $\ell_s=200$ to $\ell_s=150$ increases the signal-to-noise by about~15\%. The leakage fraction $\delta$ is less than 10\% as long as $\ell_s\gtrsim 140$.}
\label{fig:ls}
\end{figure}

The above results were derived using our canonical choice of filtering: the Gaussian filter (\ref{equ:Fell}) with $\ell_s=200$.  
Slight improvements in the signal-to-noise are possible by optimizing the smoothing scheme. Fig.~\ref{fig:ls} shows the dependence of the signal-to-noise and the leakage fraction on the smoothing parameter $\ell_s$ for $r=0.13$ and $\epsilon_L= \{\epsilon_s,\epsilon_d\}=0.5$.
We see that the signal-to-noise initially increases with~$\ell_s$, reaches a maximum at $\ell_s \simeq 120$, and then decreases as more small-scale noise is allowed for higher $\ell_s$. At the maximum, ${\rm S/N}=2.2$ and 3.8 for the Keck Array and the Simons Array, respectively. The leakage fraction at the maximum is $\delta=0.11$. 

For optimal results, we pick the smoothing scale in such a way that it maximizes the signal-to-noise while keeping $\delta<0.1$ for all values of $r$ that yield ${\rm S/N}>3$. The optimal smoothing scale for both experiments is then $\ell_s=150$, giving a 15\% increase in the signal-to-noise (see fig.~\ref{fig:ls}).\footnote{We have also tested other forms of filtering functions. 
For example, using a tanh-filter, we were able to achieve a 10 to $20\%$ improvement on the overall signal-to-noise with similar degrees of leakage for various parameters and values of $r$. 
This is because the tanh-filter is characterized by two smoothing parameters (the cut-off scale and the width), and this extra degree of freedom allows us to control the filtering process more precisely, giving us more optimized results. However, for simplicity of presentation, all the results in the paper were produced with the Gaussian filter (\ref{equ:Fell}). \label{foot:filter} } With this optimization, a more than $3\sigma$ detection becomes possible with the Simons Array even for only modest amounts of cleaning, $\epsilon_L=\{\epsilon_s,\epsilon_d\}=0.5$.
To achieve a similar level of significance with the Keck Array, we still require a high level of cleaning, $\epsilon_L=\{\epsilon_s,\epsilon_d\}=0.1$.

\begin{figure}[t]
\centering
\includegraphics[width=0.45\textwidth]{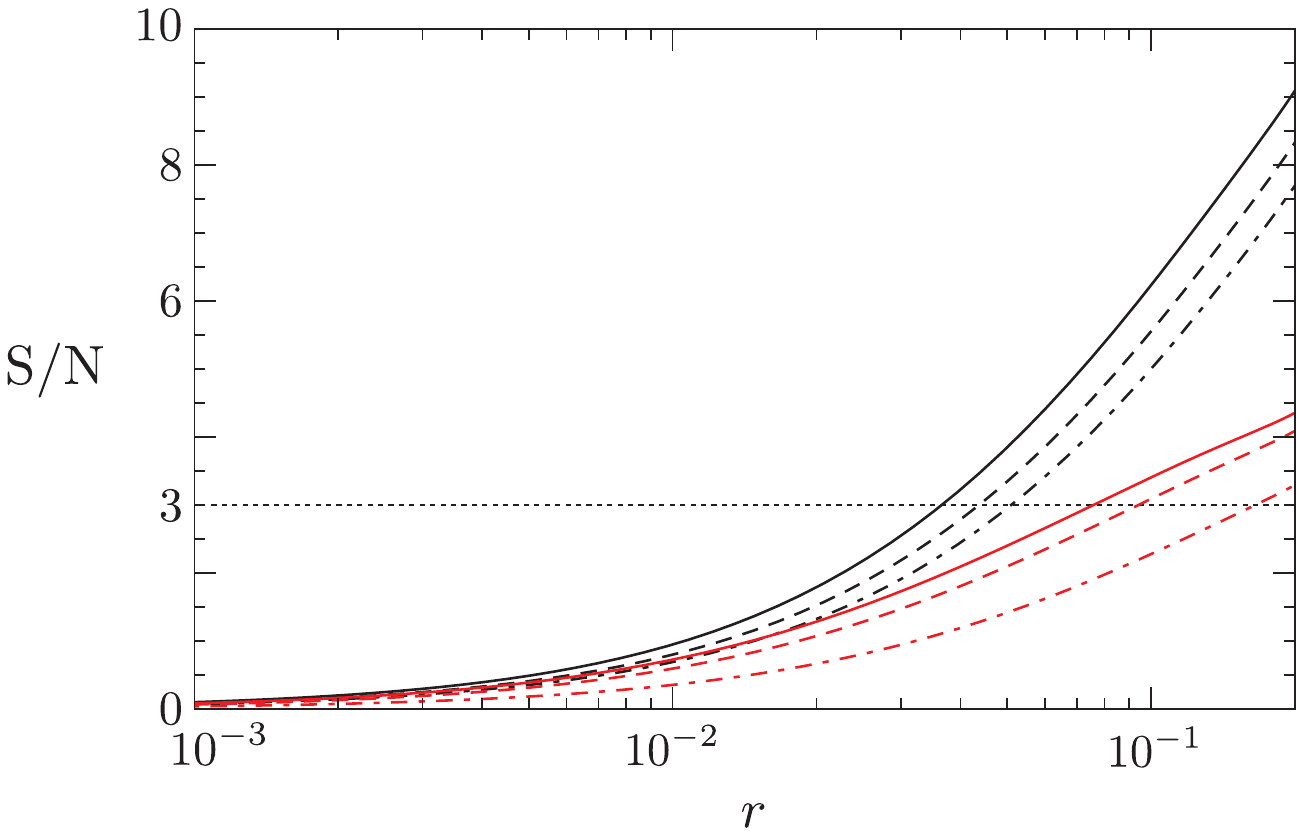}
\vskip -4pt
\caption{Signal-to-noise on the extended interval $[1.0^\circ,6.0^\circ]$ as a function of $r$ for experiments with Keck Array (red) and Simons Array (black) specifications. The foreground amplitudes have been fixed to the mean values in (\ref{As}) and (\ref{Ad}). 
The different curves correspond to $\epsilon_L=\{\epsilon_s,\epsilon_d\}=0.1$ (solid), $\epsilon_L=0.5$, $\{\epsilon_s,\epsilon_d\}=0.1$ (dashed), and $\epsilon_L=\{\epsilon_s,\epsilon_d\}=0.5$~(dot-dashed). }
\label{fig:debiased}
\end{figure}

\vskip 4pt
One may argue that we have been too conservative by choosing $\theta_{\rm min} =2.6^\circ$ as our criterion for the superhorizon signal. In particular, as can be seen from fig.~\ref{fig:CBfiltered}, a large part of the inflationary superhorizon signal isn't captured by this definition. In order to quantify the size of the total signal, we therefore also consider the extended interval with $\theta_{\rm min} = 1^\circ$. We use the debiased estimator so that the known leakage of inflationary subhorizon modes is corrected for.
Fig.~\ref{fig:debiased} shows the signal-to-noise on the interval $[1.0^\circ,6.0^\circ]$ as a function of $r$ without optimization of the filtering. We see that a $3\sigma$ detection will be possible if $r\gtrsim 0.1$ and 0.04 for the Keck Array and the Simons Array, respectively, assuming a modest amount of delensing and foreground removal of 50\%. With the optimization described above, we get ${\rm S/N}>3$ if $r\gtrsim 0.05$ and 0.025 for the Keck Array and the Simons Array, respectively.  
 While this detection wouldn't constitute a perfect causality test, it would still be a strong indication for inflationary superhorizon tensors. Moreover, at sufficiently high ${\rm S/N}$ it will be possible to measure the shape of the signal in fig.~\ref{fig:CBfiltered}, which would further strengthen this interpretation.
 
\subsection{Space-Based Experiments}
\label{sec:space}

To perform a true causality test, the $\B$-mode signal has to be measured above $\theta_{\rm min} = 2.6^\circ$. We have seen that, for $r >0.1$, this is (marginally) possible with ground-based experiments.
For $r < 0.1$, on the other hand, a future satellite mission will be required. 
For purposes of illustration, we now examine the LiteBIRD~\cite{Matsumura:2013aja} and COrE~\cite{Bouchet:2011ck} proposals. 

\vskip 4pt
All-sky surveys don't have the luxury of observing only the cleanest patches of the sky, so we need to adjust our estimates for the expected foreground levels accordingly.
The level of polarized synchrotron emission is constrained by the WMAP polarization measurements between 23 and 94 GHz~\cite{Page:2006hz}. 
Those results imply
\begin{equation}
A_s\ \simeq\ 5.8 \times 10^{-7}\, \mu{\rm K}^2 \ ,
\end{equation}
which is comparable to the 95\% upper limit of the synchrotron amplitude determined by DASI~\cite{Leitch:2004gd}.
For polarized dust emission, we take the template used by the Planck collaboration in~\cite{Aumont, Ade:2014zja} which, for $f_{\rm sky}=0.7$, gives 
\beq
A_d \ \simeq\ 5.5 \times 10^{-5}\, \mu{\rm K}^2\ .
\eeq
This choice is consistent with the FDS model~\cite{Finkbeiner:1999aq} with an average polarization fraction of about~7\%. Both of the above amplitudes are defined with respect to $\nu_0 = 100\, {\rm GHz}$ and $\ell_0=100$. 

\subsubsection{LiteBIRD}

LiteBIRD~\cite{Matsumura:2013aja} is a next-generation full-sky satellite experiment, optimized to probe large-scale B-mode polarization. It is equipped with six frequency bands in the range from 60 to 280 GHz. This frequency coverage is wide enough to perform a high level of foreground removal of both synchrotron and dust~\cite{Katayama:2011eh}.  We will therefore consider relatively small values of $\epsilon_s$ and $\epsilon_d$, namely 0.1 (realistic) and 0.01 (optimistic).
The large beams of the LiteBIRD experiment mean that delensing will only be possible in a joint analysis with external data sets~\cite{Namikawa:2014yca}.  We will assume that this will be possible only to a modest degree, $\epsilon_L \ge 0.5$.

\subsubsection{COrE}

COrE~\cite{Bouchet:2011ck} is a proposed space mission which is anticipated to deliver a full-sky CMB polarization map with a sensitivity 10 to 30 times better than its predecessor Planck. With 15 frequency bands between 45 and 795~GHz, COrE will allow a very high degree of foreground cleaning, so we will consider $\{\epsilon_s, \epsilon_d\} = $ 0.1 (pessimistic) and 0.01 (realistic).
The small beams of COrE also mean that a significant amount of internal delensing can be achieved, so we take a delensing fraction of $\epsilon_L = 0.1$ as a realistic assumption~\cite{Bouchet:2011ck}.

\subsubsection{Results}

\begin{figure}[t]
\centering
\includegraphics[width=0.45\textwidth]{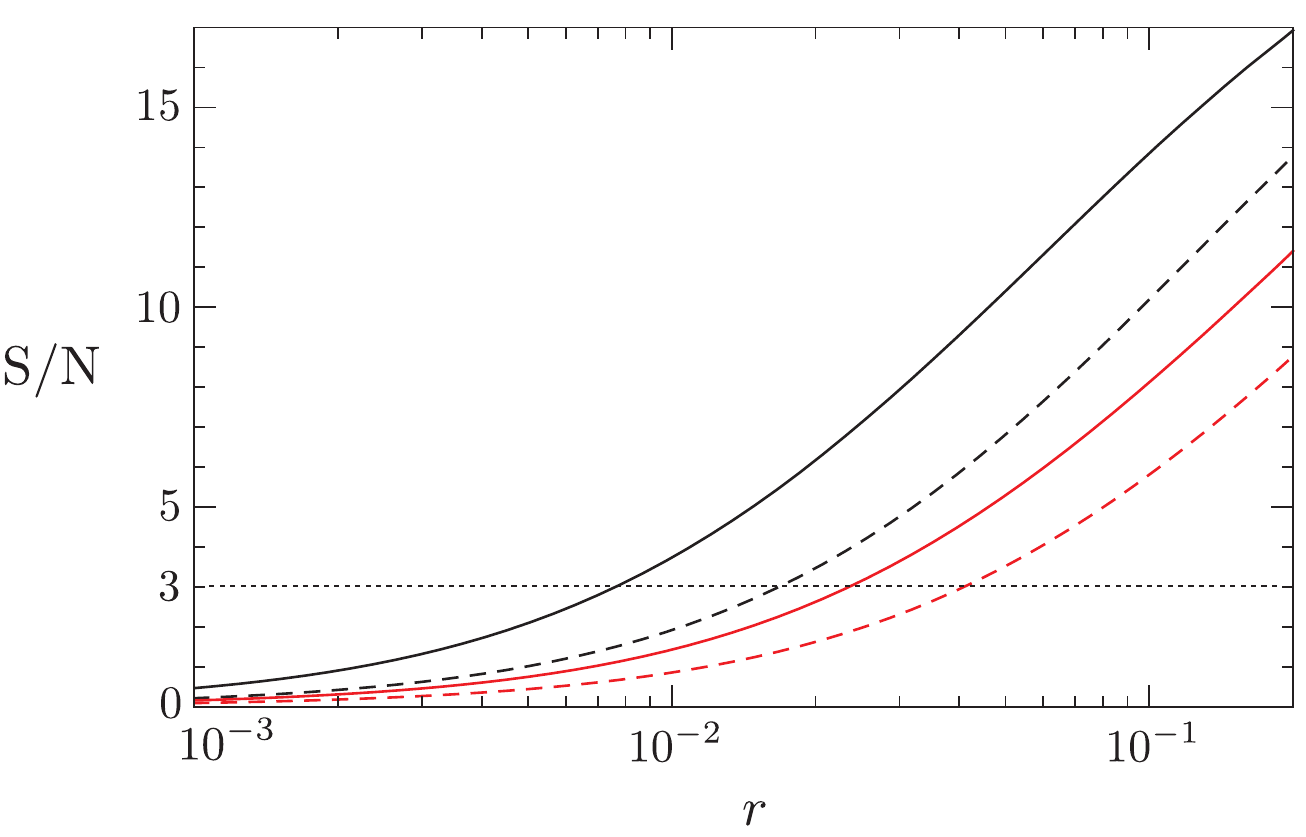}
\vskip -4pt
\caption{Signal-to-noise on the interval~$[2.6^\circ,6.0^\circ]$ as a function of $r$ for experiments with COrE (black) and LiteBIRD (red) specifications. The solid lines correspond to $\{\epsilon_s,\epsilon_d\}=0.01$, while the dashed lines assume $\{\epsilon_s,\epsilon_d\}=0.1$. The delensing fractions have been fixed to 
$\epsilon_L=0.5$ and $\epsilon_L=0.1$ for LiteBIRD and COrE, respectively.} 
\label{fig:SNR2}
\end{figure}

Fig.~\ref{fig:SNR2} displays the signal-to-noise for LiteBIRD and COrE as a function of $r$. %
We see that a $3\sigma$ detection will be possible if $r>0.04$ (0.01) with $\{\epsilon_s,\epsilon_d\}=0.1$, and $r>0.02$ (0.007) with $\{\epsilon_s,\epsilon_d\}=0.01$, for LiteBIRD~(COrE).
Depending on the actual delensing level attained by these experiments, the detection bounds stated above may shift slightly. In any case, incorporating the optimization scheme described earlier, the signal-to-noise can be improved by about 20\%. 
Thus, both LiteBIRD and COrE are capable of detecting the superhorizon $\B$-mode signal for $r\gtrsim 0.01$, in most realistic scenarios. 
For $0.001<r<0.01$, a statistically significant detection will only be possible if the extended interval $[1.0^\circ,6.0^\circ]$ is used.

\subsection{Summary}\label{sec:summary}

\begin{figure}[ht]
\centering
\includegraphics[width=0.42\textwidth]{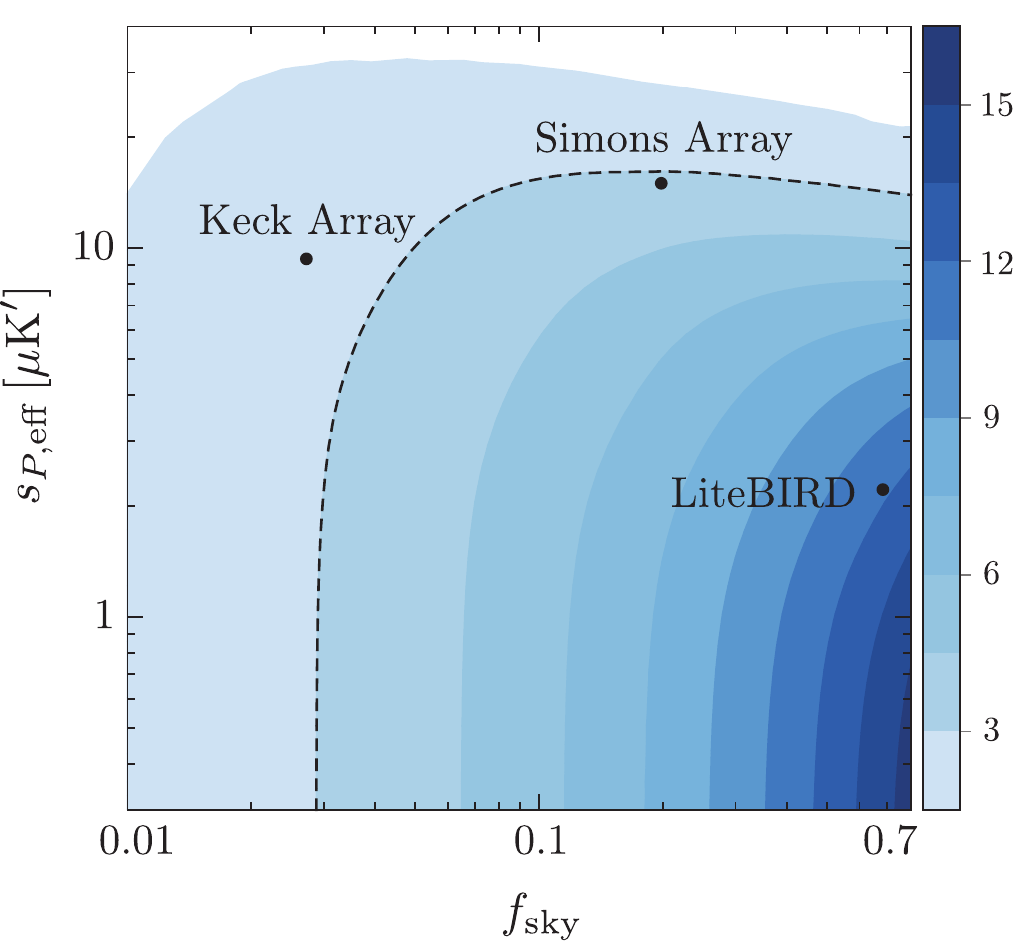}
\vskip -4pt
\caption{ Signal-to-noise on the interval $[2.6^\circ,6.0^\circ]$ for $r=0.13$ as a function of $f_\text{sky}$ and $s_{P,{\rm eff}}$.  The plot was created using the optimized Gaussian filter with $\ell_s=150$ and assumes 50\% delensing.  The dashed line indicates the $3\sigma$ detection bound.}
\label{fig:sfsky}
\end{figure}

The conclusions of this section are summarized in fig.~\ref{fig:sfsky}, which shows the signal-to-noise on the interval $[2.6^\circ,6.0^\circ]$ for $r=0.13$ as a function of the sky fraction $f_{\rm sky}$ and the effective instrumental sensitivity $s_{P,\rm eff}$. 
This time the residual foreground amplitudes have been fixed to $A_s=5.8 \times 10^{-9}\, \mu{\rm K}^2$ and $A_d=5.5\times 10^{-7}\, \mu{\rm K}^2$ at $\nu_0=100\, {\rm GHz}$, $\ell_0=100$.
As we can see, for experiments with high instrumental sensitivity, $s_{P,{\rm eff}}\lesssim 20\, \mu{\rm K}'$, sky coverage is the main factor determining whether the signal is detectable. This is because $\text{S/N}\propto \sqrt{f_\text{sky}}$ 
 in the cosmic variance limit, whereas $\text{S/N}\propto 1/s_{P,{\rm eff}}^2$ for experiments dominated by instrumental noise.
Hence, full-sky satellite missions have the best prospects for measuring the superhorizon $\B$-mode signal, though ground-based experiments such as the Simons Array can be feasible, if $r \gtrsim 0.1$. 

\section{Conclusions}\label{sec:conclusion}

The significance of a detection of primordial B-modes cannot be overstated~\cite{Baumann:2008aq, Abazajian:2013vfg, Baumann:2014nda, Baumann:2014cja}. Hence, the community is eagerly awaiting a confirmation of the cosmological character of the signal observed by the BICEP2 team~\cite{Ade:2014xna}.  However, even if the signal is established to be of primordial origin, we still wish to determine whether it was generated by vacuum fluctuations during inflation or has an alternative, post-inflationary origin.

In this paper, we have revisited the proposal of~\cite{Baumann:2009mq} for using the superhorizon part of the B-mode spectrum in real space as a model-insensitive diagnostic of inflationary gravitational waves. We found that the causality test for B-modes in its original form is not unambiguous, since we must deal with the issue of the mixing between subhorizon and superhorizon modes that is induced by the finite resolution of the experiment and the smoothing of the raw data. 
We have quantified this effect and shown how future experiments have to be designed in order to maximize the signal-to-noise of the superhorizon signal while rejecting unwanted contaminations from spurious subhorizon modes.

We have found that future ground-based experiments are capable of detecting the superhorizon $\B$-mode signal at more than $3\sigma$ significance, if the tensor-to-scalar ratio is as large as suggested by BICEP2~\cite{Ade:2014xna}, i.e.~if $r\gtrsim 0.1$. If the value of $r$ is significantly smaller, then the measurement will require a full-sky survey. 
We have found that a $3\sigma$ detection is possible with LiteBIRD and COrE as long as $r\gtrsim 0.01$, and if 90\% foreground cleaning and more than 50\% delensing can be achieved. 

We believe that using the superhorizon estimator is a powerful model-independent way to test for the inflationary origin of tensor modes and look forward to seeing it applied to future data, including the experiments considered in this work. 

\vskip 10pt
{\it Acknowledgements.} We thank Anthony~Challinor, Levon~Pogosian and Matias~Zaldarriaga for helpful discussions. D.B. and H.L.~gratefully acknowledge support from the European Research Council (ERC STG grant 279617). H.L.~also acknowledges support from the Cambridge Overseas Trust, the Lord Rutherford Memorial Research Fellowship, the Sims Empire, and the William Georgetti Scholarship. S.S.~acknowledges support from the Croucher Foundation.

\newpage
\appendix
\newpage
\section{Analysis in Harmonic Space}
\label{sec:harmonic}

The analysis in \S\ref{sec:method} and \S\ref{sec:forecast} was presented mostly in real space.  In this appendix, we give a few details of an equivalent formulation in harmonic space.

\subsection{Superhorizon Estimator}

\begin{figure}[h!]
\centering
\includegraphics[width=0.45\textwidth]{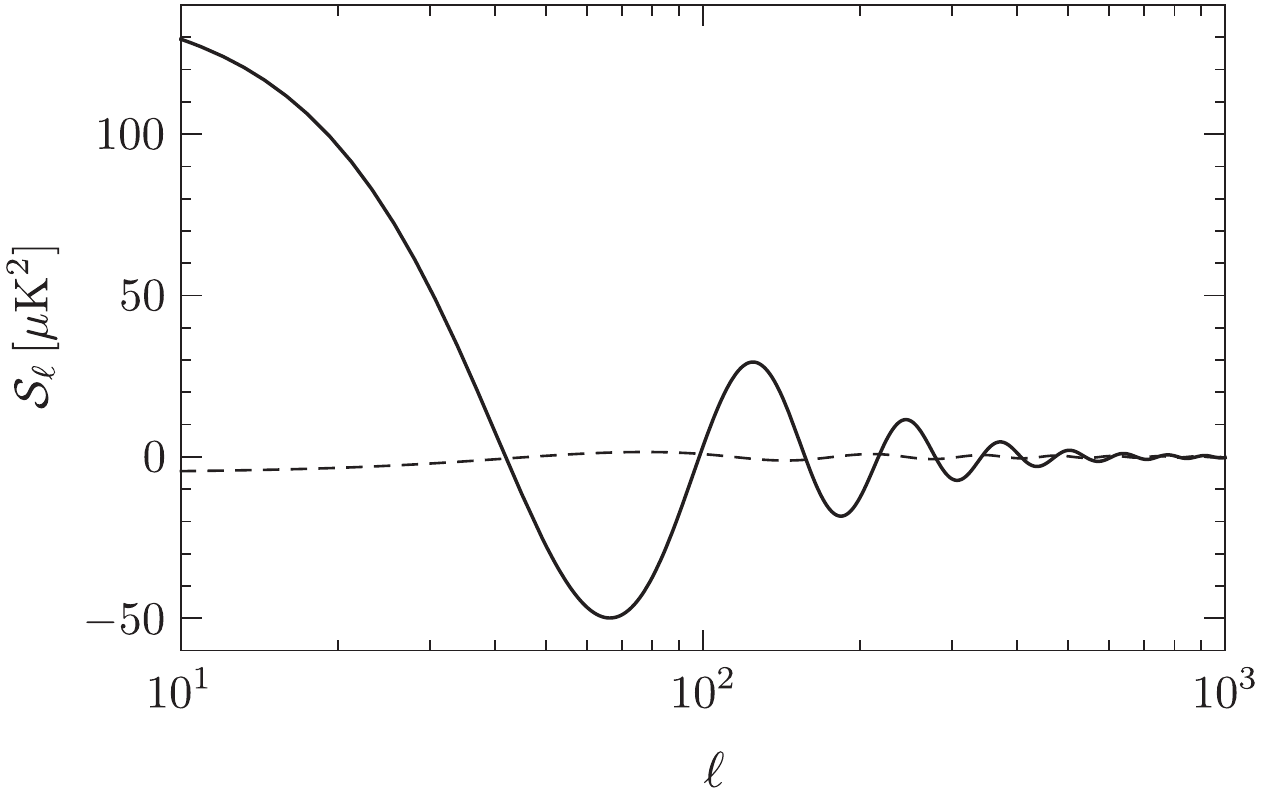}
\vskip -2pt
\caption{Local $\B$-mode power spectrum for $r=0.13$ using the Gaussian filter (\ref{equ:Fell}) with $\ell_s=200$ projected onto the interval~$[2.6^\circ, 6.0^\circ]$. The solid and dashed lines correspond to the superhorizon and subhorizon modes, respectively.} \label{fig:PSfiltered}
\end{figure}

Transforming (\ref{realestimator}) to harmonic space, we obtain an
estimator of the superhorizon part of the $\B$-mode power spectrum 
\begin{equation}\label{supestimator}
\widehat{\mathcal{S}}_\ell = \sum_{\ell'} {M}_{\ell\ell'} \widehat{\cal C}_{\ell'}\ ,
\end{equation}
where ${M}_{\ell\ell'}$ denotes a generalization of the kernel~(\ref{couplingmtx}) to the interval $\Theta = [\theta_{\rm min}, \theta_{\rm max}]$,
\begin{equation}
{M}_{\ell\ell'} \equiv \frac{2\ell'+1}{2} \underbrace{\int_{\cos \theta_{\rm max}}^{ \cos \theta_{\rm min}} P_\ell(x)P_{\ell'}(x)\hskip 1pt \d x}_{\equiv\, {I}_{\ell\ell'}}\ .
\end{equation}
The off-diagonal terms of ${I}_{\ell\ell'}$ are given by
\begin{equation}
{I}_{\ell\ell'} = \left[\frac{(\ell-\ell') x P_\ell P_{\ell'} + \ell' P_\ell P_{\ell'-1} - \ell P_{\ell-1} P_{\ell'}}{\ell(\ell+1)-\ell'(\ell'+1)} \right]_{\cos \theta_{\rm max}}^{\cos\theta_{\rm min}}\ , 
\end{equation}
while the diagonal terms still obey the recursion relation~(\ref{Idiag}). The covariance matrix of the estimator~(\ref{supestimator}) is then given by
\begin{equation}\label{ellcov}
\mathscr{C}[\widehat{\mathcal{S}}_\ell,\widehat{\mathcal{S}}_{\ell'}] = \sum_{ll'} \mathscr{C}[\widehat{\cal C}_l,\widehat{\cal C}_{l'}] \, {M}_{\ell l} {M}_{l'\ell'}\ ,
\end{equation}
where $\mathscr{C}[\widehat{\cal C}_l,\widehat{\cal C}_{l'}] $ was given in (\ref{equ:CovMat}).
Fig.~\ref{fig:PSfiltered} shows the filtered superhorizon and subhorizon $\B$-mode spectra projected onto the interval $[2.6^\circ, 6.0^\circ]$. 

\subsection{Signal-to-Noise}

We define the binned signal as
\beq
\widehat{\cal S}_b \equiv \sum_\ell B_{b\ell}\hskip 1pt \widehat{\mathcal{S}}_\ell\ ,
\eeq
where $B_{b\ell} $ is a binning matrix with uniform weight:
\begin{equation}
B_{b\ell} \equiv \left\{ \begin{array}{ll} (\ell_{(b+1)}-\ell_{(b)})^{-1} &\quad \ell_{(b)} \le \ell < \ell_{(b+1)} \\[6pt] 0 &\quad  \text{otherwise} \end{array} \right. \ .
\end{equation}
The binned covariance matrix is given by
\begin{equation}
\mathscr{C}_{bb'} \equiv \sum_{\ell\ell'} \mathscr{C}[\widehat{\mathcal{S}}_\ell,\widehat{\mathcal{S}}_{\ell'}] \, B_{b\ell} B_{\ell'b'}\ , \label{equ:CM}
\end{equation}
and the signal-to-noise is 
\begin{equation}\label{SNRell}
({\rm S/N})^2 = \sum_{bb'} \widehat{\mathcal{S}}_b \mathscr{C}^{-1}_{bb'} \widehat{\mathcal{S}}_{b'}\ ,
\end{equation}
where $\mathscr{C}^{-1}_{bb'}$ is the inverse (\ref{equ:CM}). As in the real space treatment, one has to choose the binning sensibly in order to sample the signal and the covariance well. A natural bandwidth in this case is $\Delta\ell \simeq 180^\circ/(\theta_{\rm max}-\theta_{\rm min})$.

\vskip 4pt
We have computed the signal-to-noise~(\ref{SNRell}) and compared it with the real space results quoted in $\S$\ref{sec:forecast}.
Since both treatments produce very similar results, we have chosen only to present the real space analysis in the main text.
The agreement is expected as (\ref{supestimator}) is an exact harmonic counterpart of the estimator (\ref{realestimator}). A slight difference arises from the choice of binning, since uniform binning in real space does not correspond to uniform binning in harmonic space (and vice versa). 

\section{Multi-Frequency Effective Noise}
\label{sec:EffectiveNoise}

Observations of the CMB anisotropies at multiple frequencies allow for foreground cleaning because the foreground contaminations have spectral distributions that are different from the Planck spectrum of the primordial CMB signal. In general, the different frequency channels have different noise power spectra, and the effective noise level of a multi-frequency experiment is given by taking a weighted combination which minimizes the variance~\cite{Tegmark:1999ke}. In this appendix, we assume that foreground cleaning has been performed down to a given level, and derive the effective noise for the combined foreground-cleaned CMB map. 

\vskip 4pt
The harmonic coefficients of a CMB map measured at frequencies $\nu_i$ can be written as
\begin{equation}
 a_{i,\ell m} = a^\text{CMB}_{i,\ell m} + a^\text{R}_{i, \ell m}+ a^\text{N}_{i,\ell m}\ ,
\end{equation}
where $a^\text{CMB}_{i,\ell m}$ denotes the sum of the primary CMB and the lensing-induced signal (which are both frequency-independent in thermodynamic temperature units), while $a^\text{R}_{i, \ell m}$ and $a^\text{N}_{i,\ell m}$ stand for the foreground residuals and  instrumental noise, respectively. 
We assume that the CMB signals, the foreground residuals, and the instrumental noise are uncorrelated, i.e.~for any frequency channels $i$ and $j$, we have
\begin{equation}
 \langle a_{i,\ell m}^\text{CMB} a_{j,\ell m}^\text{N}\rangle = \langle a_{i,\ell m}^\text{CMB} a_{j,\ell m}^\text{R}\rangle = \langle a_{i,\ell m}^\text{N} a_{j,\ell m}^\text{R}\rangle = 0\ .
\end{equation}
Moreover, we assume that instrumental noise at different channels are uncorrelated, so that the noise cross-power spectrum is defined as
\begin{equation}
 \langle a^\text{N}_{i,\ell m} a^{\text{N}*}_{j,\ell' m'}\rangle =  N_{i,\ell}\delta_{\ell\ell'}\delta_{mm'} \delta_{ij}\ .
\end{equation}
Although foregrounds are in general correlated among different channels---a fact that is exploited in component separation methods for foreground subtraction---we treat the foreground {\it residuals} as an extra source of {\it uncorrelated} noise. 
Thus, we have 
\begin{equation}
 \langle a^\text{R}_{i,\ell m} a^{\text{R}*}_{j,\ell' m'}\rangle =  R_{i,\ell} \delta_{\ell\ell'}\delta_{mm'} \delta_{ij}\ .
\end{equation}
Defining an estimator of the cross-power spectrum of the primary CMB as
\begin{equation}
 \widehat{C}_{ij,\ell} \equiv \sum_m \frac{a_{i,\ell m} a^*_{j,\ell m}}{2\ell+1} - L_\ell - \delta_{ij}(N_{i,\ell}+R_{i,\ell})\ ,
\end{equation}
the covariance matrix is
\begin{align}\label{CovarianceMatrixUncorrelated}
& \mathscr{C}[\widehat{C}_{ij,\ell}, \widehat{C}_{i'j',\ell'}] = \frac{\delta_{\ell\ell'}}{2\ell+1}\Big\{\big[C_\ell + L_\ell + \delta_{ii'}(N_{i,\ell} + R_{i,\ell})\big] \nonumber\\
 &\hspace{1cm}\big[C_\ell + L_\ell + \delta_{jj'}(N_{j,\ell} + R_{j,\ell})\big] + ( i' \leftrightarrow j') \Big\}\ .
\end{align}
The estimator of the power spectrum of a linearly combined foreground-cleaned map can then be expressed as
\begin{equation}
\widehat{\mathbb C}_\ell \equiv \frac{1}{Z_{\ell}}\sum_{ij}\omega_{ij,\ell}\hskip 1pt \widehat{C}_{ij,\ell}\ ,
\end{equation}
where $Z_{\ell}\equiv \sum_{ij}\omega_{ij,\ell}$. The optimal weights $\omega_{ij,\ell}$ are determined by minimizing the variance of $\widehat{\mathbb C}_\ell$. 
A straightforward computation leads to
\begin{equation}
 \omega_{ij,\ell} = \frac{1}{(N_{i,\ell}+R_{i,\ell})(N_{j,\ell}+R_{j,\ell})}\ .
\end{equation}
The minimum variance of the combined CMB map is then
\begin{equation}\label{MinVarianceUncorrelated}
 \text{Var}[\widehat{\mathbb C}_\ell]=\frac{2}{2\ell+1}\left(C_\ell + L_\ell +N^\text{eff}_\ell\right)^2\ ,
\end{equation}
where the effective noise power spectrum is defined as 
\begin{equation}
 N^{\rm eff}_{\ell} \equiv \frac{1}{\sqrt{Z_{\ell}}}= \left[\sum_{i}\frac{1}{N_{i,\ell}+R_{i,\ell}}\right]^{-1}\ .
\end{equation}
This recovers eq.~(117) of \cite{Baumann:2008aq}.

\vfil

\onecolumngrid
\newpage
\section{Experimental Specifications}
\label{sec:Experiments}

\begin{table}[h]
\centering
\begin{tabular}{ l c c r r r c c}
\hline
& $f_{\rm sky}\,[\%]$ & $\nu\, [{\rm GHz}]$ & $\theta_b\,[']$ & $N_{\rm det}$& $\Delta_{P}\,[\mu\text{K}']$ & $\Delta_{P,{\rm eff}}\,[\mu\text{K}']$ & $s_{P,{\rm eff}}\,[\mu\text{K}']$ \\ [0.5ex] 
\hline
${\rm Planck}$ &\hphantom{.}70& \hphantom{1}30 & 33\hphantom{1} & \hphantom{1}4 & 287.4\hphantom{1}\\
& & \hphantom{1}44 & 28\hphantom{1}& \hphantom{1}6& 338.9\hphantom{1}\\
& & \hphantom{1}70 & 13\hphantom{1}& 12& 298.7\hphantom{1}\\
& & 100 & 9.6\hphantom{1}& \hphantom{1}8& \hphantom{1}44.2\hphantom{1}\\
& & 143 & 7.2\hphantom{1}& \hphantom{1}8& \hphantom{1}33.3\hphantom{1}\\
& & 217 & 4.9\hphantom{1}& \hphantom{1}8& \hphantom{1}49.4\hphantom{1}\\
& & 353 & 4.7\hphantom{1}& \hphantom{1}8& 185.3\hphantom{1} & 23.0 & 27.5\\[0.5ex]
\hline
BICEP2 & 2.4 & 150 & 29\hphantom{1}& \hphantom{1}512 & \hphantom{11}5.2\hphantom{1} \\[0.5ex]
Keck Array &  &\hphantom{1}95 & 29\hphantom{1} & \hphantom{1}576 & \hphantom{11}9.0\hphantom{1}\\
 & & 150 & 29\hphantom{1} & 2560 & \hphantom{11}2.3\hphantom{1} \\
 & & 220 & 29\hphantom{1} & 1536 & \hphantom{1}10.2\hphantom{1} &  & \\[0.5ex]
BICEP3 & & \hphantom{1}95 & 29\hphantom{1} & 2560 &  \hphantom{11}2.0\hphantom{1} & \hphantom{1}1.4 & \hphantom{1}9.0\\[0.5ex]
\hline
SPTPol & \hphantom{1.}6 & \hphantom{1}90 & 1.6\hphantom{1} & 360 & 9.0\hphantom{1} \\
 & & 150 & 1.0\hphantom{1} & 1176 & 5.0\hphantom{1} & \hphantom{1}4.4 & 17.8 \\
\hline
PolarBeaR-2 & \hphantom{.}20 & \hphantom{1}95 & 5.2\hphantom{1} & 3794 &  \hphantom{1}15.1\hphantom{1} &  & \\
& & 150 & 3.5\hphantom{1} & 3794 &  \hphantom{1}15.1\hphantom{1} &  10.7 & 23.9\\[0.5ex]
\hline
Simons Array & \hphantom{.}20 & \hphantom{1}95 & 5.2\hphantom{1}& \hphantom{1}7588 & \hphantom{1}10.7\hphantom{1}\\
 & & 150 & 3.5\hphantom{1} & 11382 &  \hphantom{11}8.7\hphantom{1}\\
 & & 220 & 2.7\hphantom{1} & \hphantom{1}3794 &  \hphantom{1}16.7\hphantom{1} & \hphantom{1}6.3 & 14.1\\[0.5ex]
\hline
LiteBIRD & \hphantom{.}70 & \hphantom{1}60 & 32\hphantom{1} & \hphantom{1}304 &  \hphantom{1}10.3\hphantom{1}\\ 
 & & \hphantom{1}78 & 58\hphantom{1} & \hphantom{1}304 &  \hphantom{11}6.5\hphantom{1} \\
 & & 100 & 45\hphantom{1} & \hphantom{1}304 &  \hphantom{11}4.7\hphantom{1} \\
 & & 140 & 32\hphantom{1} & \hphantom{1}370 &  \hphantom{11}3.7\hphantom{1} \\
 & & 195 & 24\hphantom{1} & \hphantom{1}370 &  \hphantom{11}3.1\hphantom{1} \\
 & & 280 & 16\hphantom{1} & \hphantom{1}370 &  \hphantom{11}3.8\hphantom{1} & \hphantom{1}1.8 & \hphantom{1}2.2\\ [0.5ex]
\hline
COrE & \hphantom{.}70 & \hphantom{1}45& 23.3\hphantom{1} & \hphantom{11}64 &  \hphantom{1111}9.1\hphantom{1} \\
& & \hphantom{1}75 & 14.0\hphantom{1} & \hphantom{1}300 &  \hphantom{1111}4.7\hphantom{1}\\
& & 105 & 10.0\hphantom{1} & \hphantom{1}400 &  \hphantom{1111}4.6\hphantom{1}\\
& & 135 & 7.8\hphantom{1} & \hphantom{1}550 &  \hphantom{1111}4.6\hphantom{1}\\
& & 165 & 6.4\hphantom{1} & \hphantom{1}750 &  \hphantom{1111}4.6\hphantom{1}\\
& & 195 & 5.4\hphantom{1} & 1150 &  \hphantom{1111}4.5\hphantom{1}\\
& & 225 & 4.7\hphantom{1} & 1800 &  \hphantom{1111}4.6\hphantom{1}\\
& & 255 & 4.1\hphantom{1} & \hphantom{1}575 &   \hphantom{111}10.5\hphantom{1}\\
& & 285 & 3.7\hphantom{1} & \hphantom{1}375 &  \hphantom{111}17.4\hphantom{1}\\
& & 315 & 3.3\hphantom{1} & \hphantom{1}100 &  \hphantom{111}46.6\hphantom{1}\\
& & 375 & 2.8\hphantom{1} & \hphantom{11}64 &  \hphantom{111}19.0\hphantom{1}\\
& & 435 & 2.4\hphantom{1} & \hphantom{11}64 &  \hphantom{11}258.0\hphantom{1}\\
& & 555 & 1.9\hphantom{1} & \hphantom{11}64 & \hphantom{11}626.0\hphantom{1}\\  
& & 675 & 1.6\hphantom{1} & \hphantom{11}64 & \hphantom{1}3640.0\hphantom{1}\\
& & 795 & 1.3\hphantom{1} & \hphantom{11}64 & 22200.0\hphantom{1} & \hphantom{1}1.8 & \hphantom{1}2.2\\[1ex] 
\hline\hline
\end{tabular} 
\caption{Instrumental specifications for current and planned CMB polarization experiments \cite{Ade:2014gua, Bock, Austermann:2012ga, Lee, Matsumura:2013aja, Bouchet:2011ck}.}
\label{tab:specs2}
\end{table}

\vfil

\newpage
\bibliography{References}
\end{document}